\documentclass[12pt]{iopart}
\usepackage{iopams}

\usepackage{graphicx}
\usepackage{dcolumn}
\usepackage{bm}
\usepackage{mathbbol}
\usepackage{epsf}
\usepackage{mathrsfs}
\usepackage{slashed}
\usepackage{stmaryrd}

\newcommand{\be}{\begin{equation}}
\newcommand{\ee}{\end{equation}}
\newcommand{\bea}{\begin{eqnarray}}
\newcommand{\eea}{\end{eqnarray}}
\newcommand{\bml}{\begin{mathletters} \baselineskip 10pt}
\newcommand{\eml}{\baselineskip 12pt \end{mathletters}}
\newcommand{\nn}{\nonumber}

\newcommand{\m}{{\scriptscriptstyle -}}
\newcommand{\p}{{\scriptscriptstyle +}}

\newcommand{\PP}{\mathbb{P}}

\newcommand{\ud}{\mathrm{d}}
\newcommand{\pathD}{\!\mathscr{D}}
\newcommand{\x}{\mathbf{x}}

\def\lambdabar{\protect\@lambdabar}
\def\@lambdabar{%
\relax \bgroup
\def\@tempa{\hbox{\raise.73\ht0
\hbox to0pt{\kern.2\wd0\vrule width.7\wd0
height.1pt depth.1pt\hss}\box0}}%
\mathchoice{\setbox0\hbox{$\displaystyle\lambda$}\@tempa}%
{\setbox0\hbox{$\textstyle\lambda$}\@tempa}%
{\setbox0\hbox{$\scriptstyle\lambda$}\@tempa}%
{\setbox0\hbox{$\scriptscriptstyle\lambda$}\@tempa}%
\egroup }

\newcommand{\pad}[2]{\frac{\partial #1}{\partial #2}}

\newcommand{\vc}[1]{\mbox{\boldmath$#1$}}
\newcommand{\svc}[1]{\mbox{\footnotesize\boldmath$#1$}}
\newcommand{\ssvc}[1]{\mbox{\scriptsize\boldmath$#1$}}

\newcommand{\bracket}[2]{\bra{#1}\,#2\rangle}
\newcommand{\bra}[1]{\langle\,#1\,|}
\newcommand{\ket}[1]{|\,#1\,\rangle}

\newcommand{\twomatrix}[1]{\bigg(\begin{array}{cc} #1 \end{array}\bigg)}

\newcommand{\FP}{\mbox{FP}}

\begin{document}

\title{Noncommutativity from spectral flow}

\author{Thomas Heinzl and Anton Ilderton}

\address{School of Mathematics and Statistics, University of
Plymouth\\
Drake Circus, Plymouth PL4 8AA, UK}

\ead{\mailto{theinzl@plymouth.ac.uk}}
\ead{\mailto{abilderton@plymouth.ac.uk}}

\begin{abstract}
We investigate the transition from second to first order systems.
This transforms configuration space into phase space and hence
introduces noncommutativity in the former. Quantum mechanically,
the transition may be described in terms of spectral flow. Gaps in
the energy or mass spectrum may become large which effectively
truncates the available state space. Using both operator and path
integral languages we explicitly discuss examples in quantum
mechanics, (light-front) quantum field theory and string theory.
\end{abstract}


\maketitle

\section{Introduction}\label{sec:1}

The last decade has seen a renaissance of the old idea of
noncommutative (`quantised') space-time \cite{Snyder:1946qz},
triggered by its reappearance in the context of string and
M(atrix) theory
\cite{Connes:1997cr,Douglas:1997fm,Schomerus:1999ug,Seiberg:1999vs}.
Applications (and publications) are numerous as is well documented
by the recent reports and texts on noncommutative geometry
\cite{Connes:1994yd,Madore:2000aq}, deformation quantisation
\cite{Dito:2002dr}, noncommutative field theory
\cite{Douglas:2001ba,Szabo:2001kg} and possible phenomenological
consequences \cite{Hinchliffe:2002km} which include noncommutative
approaches to gravity \cite{Aschieri:2005yw,Aschieri:2005zs}, the
standard model \cite{Calmet:2001na}, Lorentz violation
\cite{Carroll:2001ws} and the quantum Hall effect
\cite{Susskind:2001fb,Hellerman:2001rj}.

The latter is based on the quantum mechanics of a particle in a
plane pierced by a strong magnetic field. Similar to the string
theory scenario it is the presence of the magnetic field that
entails noncommutativity between the co-ordinates in the plane.
This physics and its exposition in the papers
\cite{Dunne:1989hv,Dunne:1992ew,Guralnik:2001ax,Jackiw:2001dj,Jackiw:2002wd}
are the original inspiration for the present work. Our main
focus is the description of noncommutativity as an
\textit{emergent phenomenon} in terms of \textit{spectral flow}.

In particular, we analyse how commutative spaces become
noncommutative in special limits of quantum mechanical theories.
The limits to be studied appear initially to be unrelated.
However, we will unveil that there are features common to all, and
indeed that there is a unifying picture.

Let us outline our approach, using a generic (spectral flow)
parameter $\lambda$ to characterise the limits in question as
$\lambda \to 0$. From an action principle point of view, the
limits we consider correspond to terms quadratic in time
derivatives (`velocities') vanishing or being rendered negligible
compared to other terms. As $\lambda \to 0$ the remaining terms
are at most linear in time derivatives, so that we move from a
second to a first order system. This implies a significant
alteration to the theory, as the definition of the conjugate
momenta and therefore the Poisson brackets of the theory will be
quite different to when $\lambda \not= 0$.

When we quantise, Poisson brackets are replaced by commutators of
operators. In an operator picture we observe that the energy
spectrum is $\lambda$--dependent. As we take the limit some
portion of states becomes highly excited and decouples from the
theory --- effectively this spectral flow truncates the available
state space. Operators which commute at $\lambda\not=0$ do so
because of cancellations between the various modes of the
operators (think of working with the Fourier modes of a scalar
field). As some of these modes are decoupled at $\lambda=0$ such
cancellations are incomplete and operators may fail to commute.
Typically, it is the configuration space (spacetime co-ordinate or
field configuration) operators which become noncommutative in the
limit, as the spectral flow takes us to a theory where
configuration space becomes phase space.

We will also investigate the limit from a functional perspective,
where we show that functionals such as the vacuum and transition
amplitudes have natural interpretations as $\lambda \to 0$ in
terms of functionals in the first order theory. Here one must
carefully take into account changes to the true degrees of freedom
(the arguments of functionals) which occur because of the shift
from configuration to phase space.

This paper is organised as follows. In section \ref{QM-sect} we
discuss the quantum mechanics of a particle in a magnetic field.
In the limit in which the magnetic field is large compared to the
particle mass we observe that co-ordinates become noncommutative.
We describe this limit in the operator language as a projection
onto the lowest energy level. From a functional perspective we
show how to take account of the change from configuration space to
phase space, giving explicit examples of the first order limit of
second order transition amplitudes. We conclude this section with
a discussion of an analogous situation in string theory, where the
presence of a strong `magnetic' field leads to an effective lower
dimensional noncommutative field theory.

In section \ref{NR-sect} we study the nonrelativistic limit of a
quantum field theory, in which states of high energy and momentum
are decoupled. We see that the second order Klein-Gordon equation
becomes the first order Schr\"odinger equation and show that only
particle-number conserving interactions survive the
non-relativistic limit.

In section \ref{LF-sect} we describe light-front quantum field
theory as a limiting transition to quantising on null-planes.
Using `almost' light-front co-ordinates we describe the energy
spectrum and show that half of the mass shell energies are
decoupled in the light-front limit. We then show explicitly how
this leads to non-zero commutators of the field with itself, and
describe the vacuum functional and time evolution generator in the
light-front limit.

We present our conclusion in section \ref{Conc-sect}. The
appendices contain some review material on relevant functional
integrals.

\section{Particle in a strong magnetic field}\label{QM-sect}

\subsection{Operator approach}

Consider a nonrelativistic particle moving in the $xy$ plane under
the influence of a constant magnetic field of magnitude $B$ in the
$z$-direction. Upon quantisation this is the problem
originally solved by Landau in 1930 \cite{landau:1930}. The
system is described by the standard Lagrangian \cite{landau:1995}
\be \label{LANDAU}
  L = \frac{m}{2} \bigg( \dot{x}^2 + \dot{y}^2 + \dot{z}^2 \bigg)
  + B \dot{x} y \; .
\ee
The conjugate momenta are
\bea
  p_x &=& m \dot x + By \equiv \pi_x + By \; , \nn \\
  p_y &=& m \dot y \equiv \pi_y\; , \quad  p_z=m \dot{z} \equiv \pi_z \; ,
\eea
and obey the equations of motion
\be
  \dot{p}_x = 0 \; , \quad \dot{p}_y = B \dot{x} \; , \quad \dot{p}_z = 0 \; .
\ee
These imply three conserved quantities $p_z$, $x_0$ and $y_0$ with
the latter two given by
\be \label{XY0}
  y_0 = y + \frac{\dot{x}}{\omega} \; , \quad x_0 = x -
  \frac{\dot{y}}{\omega} \; ,
\ee
where $\omega = B/m$ is the usual cyclotron frequency. The
relevance of these two operators in the quantum theory was first
noted by Johnson and Lippmann \cite{johnson:1949}.

Upon performing a Legendre transformation the Hamiltonian is found
to be
\be
  H = p_x \dot x + p_y \dot y - L = \frac{p_y^2}{2m}  +
  \frac{1}{2} m \omega^2 (y-y_0)^2 + \frac{p_z^2}{2m}  \; .
\ee
We introduce ladder operators
\bea
  a &=& \left( \frac{m\omega}{2} \right)^{1/2} \bigg(y-y_0 +
  \frac{i}{m\omega} p_y\bigg)\; , \\
  a^\dagger &=& \left( \frac{m \omega}{2} \right)^{1/2} \bigg(y -
  y_0 - \frac{i}{m \omega} p_y\bigg) \; ,
\eea
in terms of which the Hamiltonian may be written as
\be
 H = \omega \bigg( a^\dagger a + \frac{1}{2} \bigg) + \frac{p_z^2}{2m}  \; .
\ee
Obviously this represents a harmonic oscillator in $y$ shifted by
$y_0$, as in (\ref{XY0}), and free motion in $z$.

As operators, the conserved quantities (\ref{XY0}) commute with
the Hamiltonian and the kinematical (not conjugate) momenta. Their
commutators with the co-ordinate operators are
\bea
  ~[\hat x_0, \hat y_0] &=& iB^{-1} \; , \label{X0Y0COMM} \\
  ~[\hat x_0,\hat y] &=& [\hat x, \hat y_0] = iB^{-1} \; ,  \\
  ~[\hat x_0,\hat x] &=& [\hat y_0, \hat y] = 0  \; .
\eea
Eigenstates of the Hamiltonian are labelled by the oscillator
(Landau) level $n$ and $p_z$ but are infinitely degenerate with
respect to $p_x$,
\be
  \hat{H} \ket{n,p_x,p_z} = E_n(p_z) \ket{n,p_x,p_z} \; ,
\ee
as the energy is independent of $p_x$,
\be
  E_n (p_z) \equiv \omega \bigg( n + \frac{1}{2} \bigg) +
  \frac{p_z^2}{2m}  \; .
\ee
Note that the level spacing $\omega = B/m$ becomes large for $B
\gg m$. In this case one expects that transitions between Landau
levels are strongly suppressed and that any dynamics will be
restricted to the lowest level, $n=0$
\cite{Dunne:1989hv,Guralnik:2001ax,Jackiw:2002wd}. The projection
onto the latter is given by the operator
\be
  \mathbb{P} = \int \frac{\ud p_z \ud p_x}{(2\pi)^2} \,
  \ket{0,p_x,p_z} \bra{0,p_x,p_z} \; .
\ee
In what follows we evaluate the projected commutator, $\big[ \PP
\hat{x} \PP , \PP \hat{y} \PP \big]$ and show that it is
nonvanishing. As a preparation we note
\be
  \PP \hat{y} \PP = \PP\bigg[ \hat{y}_0 +
  \frac{1}{\sqrt{2m\omega}}(\hat a+{\hat a}^\dagger)\bigg] \PP = \PP \hat{y}_0\PP\,.
\ee
The second equality follows from $\hat a\PP = 0$ since $\PP \sim \ket{0,p_x,p_z}$ and $\PP
a^\dagger \PP = 0$ since $\PP {\hat a}^\dagger \PP \sim
\bracket{p'_z,p'_x,0}{1,p_x,p_z}=0$. Similarly,
\be
  \PP \hat{x} \PP = \PP\bigg[ \hat{x}_0 +
  \frac{1}{i\sqrt{2m\omega}}(\hat a-{\hat a}^\dagger)\bigg]\PP = \PP \hat{x}_0
\PP. \ee
Now, $\hat{y}_0$ commutes with the Hamiltonian, $\hat p_x$ and $\hat p_z$,
so
\be
    \PP\hat{y}_0\PP = \PP\hat{y}_0=\hat{y}_0\PP,
\ee
using $\PP^2=\PP$. So, finally,
\be
    \big[ \PP \hat{x} \PP , \PP \hat{y} \PP \big] = \big[ \PP
    \hat{x}_0 \PP , \PP \hat{y}_0 \PP \big]
    = \PP[\hat x_0,\hat y_0]\PP
    = i B^{-1} \, \PP \; . \label{PXYP}
\ee
We see that the projection onto the lowest energy level results in
a non-zero commutator between the (projected) position operators.

The commutator may be explained in a simple fashion
\cite{Dunne:1989hv,Guralnik:2001ax,Jackiw:2002wd} by performing
the limit $m/B \to 0$ (small mass/large field) in the Lagrangian
(\ref{LANDAU}). To retain a nontrivial theory we add an arbitrary
potential $V(x,y)$ and arrive at the \textit{first order}
Lagrangian
\be \label{LLL}
  L = B \dot{x} y - V(x,y) \; .
\ee
The Poisson bracket or commutator is read off from the first term
(the `canonical one-form' \cite{Faddeev:1988qp}) which yields
\be\label{B}
  [x,y] = iB^{-1} \; .
\ee
Hence, the configuration space variables $x$ and $y$ become a
canonical pair and thus define a phase space on which one has the
Hamiltonian
\be
  h(x,y) \equiv V(x,y) \; ,
\ee
with $By$ playing the role of the momentum conjugate to $x$
(Peierls' substitution \cite{Peierls:1933,Duval:2000xr}). The
Hilbert space of states may be taken to be $\mathbb{L}^2
(\mathbb{R})$ consisting of wave functions $\Psi = \Psi(x)$. The
emerging picture will be the basis of the following subsection.

\subsection{Path integral approach}
We have seen that at a classical level a first order theory is
obtained simply by deleting the kinetic term in the Lagrangian of
the second order theory, though quantum mechanically the limit is
somewhat more subtle. In this subsection we will study the limit
from a functional viewpoint.

We begin with a typical wave function in the second order theory,
say the amplitude describing particle transition from $(x_i,y_i)$
to $(x_f,y_f)$ in time $t=T$. In the Euclidean path integral
language this amplitude is the sum over all paths between the two
points weighted with the exponent of the classical (Euclidean)
action,
\be\label{1st-new}
  \bra{x_f,y_f}e^{- \hat{H}T}\ket{x_i,y_i} = \int\pathD
  y\!\int\pathD x\,\,\exp\, - \int\limits_0^T\!\ud t\
  L\bigg|_{x(0)=x_i,\, y(0)=y_i}^{x(T)=x_f,\, y(T)=y_f}\,,
\ee
with $L$ the Euclidean version of (\ref{LANDAU}) together with a potential $V(x,y)$.
For a review of this and similar constructions see appendices A
and B. For clarity we will neglect any $z$ dependence. Let $\hat
H$ and $E_0$ ($\hat h$ and $\epsilon_0$) be the Hamiltonian and
vacuum energy in the second (first) order theory. In this section
we will discuss the following operation and show that it gives
first order transition amplitudes in the limit of small mass/large
$B$,
\be \label{AMP21}
  \lim_{m \to 0}\, \int\! \ud y_f \int\!\ud y_i \, \bra{x_f , y_f} e^{ -
  (\hat{H} - E_0) T} \ket{x_i, y_i} =\bra{x_f} e^{-(\hat{h}-\epsilon_0)T} \ket{x_i}\, .
\ee
There are two points to consider. The first is the subtraction of
the vacuum energy and the second the integration over
\textit{half} of the boundary degrees of freedom in the second
order theory. In the operator formalism it has been seen that the
energy spectrum undergoes a flow which decouples excited states as
$m/B \to 0$. In the case of a particle in a magnetic field this
constrains the particle to lie in the ground state, that is the
lowest Landau level. Strictly, however, even the ground state of
this system acquires a divergent energy, ($\omega/2 = B/2m \to
\infty$ in the case of $V=0$), and this should be subtracted from
the Hamiltonian in order to arrive at a meaningful system. This is
most clearly seen using the spectral decomposition of the
transition amplitude,
\be
  \bra{x_f,y_f} e^{-\hat{H}T} \ket{x_i,y_i} = \sum_{n \geq 0,d}
\psi_{n,d}^* (x_f,y_f) e^{-E_n T} \psi_{n,d} (x_i,y_i) \; , \ee
where the sum (which represents any combination of discrete and
continuous measure) is over all energy eigenvalues $E_n$ of the
Hamiltonian and degeneracies $d \equiv (p_x , p_z)$ of those
energy levels \cite{Jackiw:2002wd}. If the eigenvalues flow to
infinity as the mass decreases we see that all terms in this
series are exponentially damped and in the massless limit this
expression is null. If, however, we subtract the vacuum energy
from the Hamiltonian then this sum becomes
\bea\label{specdec}
  \bra{x_f,y_f} e^{-(\hat{H} - E_0) T} \ket{x_i,y_i} &=&
  \sum_{d} \psi_{0,d}^* (x_f,y_f) \psi_{0,d} (x_i,y_i) \nn \\
  &+& \sum_{n \geq1 , d} \psi_{n,d}^* (x_f,y_f) e^{- \Delta
  E_n T} \psi_{n,d} (x_i,y_i) \; ,
\eea
where $\Delta E_n \equiv E_n-E_0$. The first term now survives the
limit, the caveat being that these manipulations are only well
defined in Euclidean space.

Moving on to the second point we consider the integration over
half of the degrees of freedom on the `boundary' $t=0$ and $t=T$.
This has a natural interpretation in the second order theory ---
rather than consider the amplitude for a transition between points
$(x_i,y_i)$ and $(x_f,y_f)$ we instead ask for the amplitude for
transition between points $x_i$ and $x_f$ for any initial and
final values of $y$. Now, as already stated, in the massless limit
the operators $\hat x$ and $\hat y$ form a conjugate pair. In
taking the limit from the second order theory, where these
variables are independent, we must choose which of $x$ and $y$ to
consider as a co-ordinate and which a momentum. In the first order
theory our prescription corresponds to a particular choice of
polarisation or (Schr\"odinger) representation, namely that where
we diagonalise the operator $\hat{x}$.

In general we must choose one linear combination of $\hat{x}$ and
$\hat{y}$ to be diagonalised and compute the transition amplitude
between eigenstates of these operators. The remaining degrees of
freedom should be integrated over at the boundary. In this way the
configuration space path integral becomes a phase space path
integral in the first order theory.  If we choose to represent
states in the massless limit by wave functions $\Psi(x)$, then we
integrate over all possible values of the momentum $y_f$ and $y_i$
at the boundaries on the left-hand side of (\ref{AMP21}) (or the
right-hand side of (\ref{1st-new})) to arrive at
\be\label{1st}
  \bra{x_f}e^{-i\hat{h}T}\ket{x_i} = \int\pathD
  y\!\int\pathD x\,\,\exp\, i\int\limits_0^T\!\Big[Bx\dot{y} -
  h(x,y)\Big]\bigg|_{x(0)=x_i}^{x(T)=x_f}\,,
\ee
which, upon changing variables $p=By$, is a standard phase space
path integral describing a transition amplitude in the lower
dimensional first order theory. We will give explicit examples
below.

\subsection{Example one: no external potential}

A simple example of the above is given by the particle in a
magnetic field with no external potential, for which the transition amplitudes
in the first order theory are simply
\be\label{simple}
    \bra{x_f}e^{-\hat h T}\ket{x_i}|_{h=0}=\delta(x_f-x_i),
\ee
We would like to relate this amplitude to the
`projected' transition amplitude in the $m\not=0$ theory,
\be\label{proj-trans1}
  \bra{x_f,y_f} \PP e^{-(\hat{H}-E_0)T}\ket{x_i,y_i} \,.
\ee
We label energy eigenstates by the oscillator level $n$ and an
eigenvalue of $\hat p_x=B\hat y_0$. The projection operator is
then written as an integral over the degenerate ground states
$\ket{0,p_x}$,
\be
  \nn \PP =\int\limits\!\frac{\ud p_x}{2\pi}\,\,\ket{0,p_x}\bra{p_x,0},
\ee
normalised such that $\bracket{0, p_x}{q_x ,0} = 2\pi \delta(p_x -
q_x)$ and so $\PP^2=\PP$. The projected amplitude
(\ref{proj-trans1}) may then be written
\bea\label{proj-trans2}\label{30}
  \bra{x_f,y_f}\PP e^{-(\hat{H}-E_0)T}\ket{x_i,y_i}&=&
  \int\limits\!\frac{\ud p_x}{2\pi}\,\,
  \bracket{x_f,y_f}{0,p_x}\bracket{0,p_x}{x_i,y_i}\,,\\
  &=& \int\limits\!\frac{\ud p_x}{2\pi}\,\,
  \psi_{0,p_x}(x_f,y_f)\,\psi^*_{0,p_x}(x_i,y_i).
\eea
Comparing this with the spectral decomposition of the full
transition amplitude (\ref{specdec}) we see that the two coincide at
large times (if we rotate to Euclidean space). The effect of the
projection is to restrict intermediate states of the system to the
ground state, so that transition amplitudes do not obtain
contributions from excited states.

To calculate the projected amplitude we first derive the explicit
form of the ground state wave function. In the above
representation this is given by the two conditions
\be
  \eqalign{
  \hat{p}_x\,\psi_{0,p}(x,y) &\equiv -i\partial_x\psi_{0,p}(x,y) =
  p\,\psi_{0,p}(x,y), \\
  \hat{a}\,\psi_{0,p}(x,y) &\equiv \bigg(\frac{B}{2m}\bigg)^{1/2}
  \bigg(y - \frac{p}{B} +
  \frac{1}{B}\partial_y\bigg)\psi_{0,p}(x,y)=0 \; ,
  }
\ee
which are respectively plane wave and harmonic oscillator equations. They have the solutions
\be
  \psi_{0,p}(x,y) = N_p\,\exp \, \left\{ ixp - \frac{B}{2}
  (y-pB^{-1})^2 \right\} \; , \qquad N_p^2 = \sqrt{B/\pi}.
\ee
Using these functions we follow the procedure of
the previous subsection, integrating (\ref{30}) over $y_f$ and $y_i$, c.f. (\ref{AMP21}),
\bea
  \int\limits\!\ud y_f\ud y_i\,\,\bra{x_f,y_f}\PP
  e^{-(\hat H-E_0)T}\ket{x_i,y_i} =
  2\,\sqrt{\frac{\pi}{B}}\,\,\delta(x_f-x_i),
\eea
recovering, up to an irrelevant normalisation effect, the trivial
transition amplitude (\ref{simple}) of the first order theory.

\subsection{Example two: harmonic oscillator potential}
We now give a non-trivial example of the massless limit. Consider the (Euclidean) action,
\begin{equation}
  S_\mathcal{E}(m) = \int\limits_0^T \! \ud t \,\, \frac{m}{2}
  (\dot{x}^2+\dot{y}^2) - iB\dot{x}y + \frac{\lambda^2}{2}\big(x^2 +
  y^2\big).
\end{equation}
Setting $m=0$ we arrive at the Euclidean action of the harmonic
oscillator (upon changing variables $p=By$) with frequency
$\mu=\lambda/B$ and mass $B^2/\lambda$. The phase space integral
which quantises the $m=0$ action is
\be
    \int\pathD x\mathscr{D}p\,\,\exp -S_\mathcal{E}(0)\bigg|_{x(0)=x_i}^{x(T)=x_f},
\ee
where the momentum $p=By$ has a free boundary. This result is well
known and in this section we will recover it from the massless
limit of the quantised second order theory.

For $m\not=0$, following (\ref{AMP21}), we integrate over boundary
values of $y$ in the transition amplitude,
\be\fl
  \int\!\ud y_i\ud y_f\bra{x_f,y_f}e^{-i\hat{H}T}\ket{x_i,y_i}
  \equiv \int\!\ud y_i\ud y_f \int\pathD y \!\int \pathD x \,\, \exp
  -S_\mathcal{E}(m)\bigg|_{x(0) = x_i, \, y(0)=y_i}^{x(T)=x_f,\,
  y(T)=y_f}.
\ee
The integrals are Gaussian and easily computed, the result is
\begin{equation}
  \frac{1}{\det^{1/2}}\,\,\exp\bigg[a_0(m) +
  \sum\limits_{n=1}^\infty a_n(m) \bigg]\,,
\end{equation}
where the determinant factor is given by
\begin{equation}
  \det \equiv \prod\limits_{n=1}^\infty \mathcal{K}_n(m) \quad
  \mathrm{with} \quad \mathcal{K}_n(m) = \bigg(
  \lambda+\frac{mn^2\pi^2}{T^2}\bigg)^2+\frac{B^2n^2\pi^2}{T^2},
\end{equation}
and the exponentiated terms are
\bea
  a_0(m) &= -\frac{T\lambda}{6}(x_f^2 + x_i^2+x_fx_i) -
  \frac{m}{2T} (x_f-x_i)^2 - \frac{B^2}{2\lambda T}(x_f-x_i)^2, \\
  a_n(m) &= \frac{\lambda^2T}{n^2\pi^2} ((-)^n x_f - x_i)^2
  \bigg( \lambda + \frac{m n^2\pi^2}{T^2} \bigg)
  \mathcal{K}_n(m)^{-1}.
\eea
The sum over $a_n(m)$ behaves as $n^{-4}$ for $n$ large and is
convergent, although the result is an unenlightening combination of
hypergeometric functions which nevertheless gives the expected
result as $m \to 0$. Rather than detail this we illustrate it by
performing the (in this case) equivalent operations of taking
$m\to0$ and then performing the sum,
\begin{equation}
  \fl    \sum\limits_{n=1}^\infty a_n(0) =(x_f^2 + x_i^2) \bigg[
  \frac{\lambda T}{6} + \frac{B^2}{2\lambda T} - \frac{B\cosh(\mu
  T)}{2\sinh(\mu T)} \bigg] - x_f x_i \bigg[ \frac{B^2}{\lambda T}
  - \frac{\lambda T}{6} - \frac{B}{\sinh(\mu T)} \bigg].
\end{equation}
Therefore, in the massless limit we find the exponential of
\begin{equation}
  \fl    a_0(0)+\lim_{m\to0}\sum a_n(m) = -\frac{B}{2\sinh(\mu
  T)}\bigg(x_f^2\cosh(\mu T)+x_i^2\cosh(\mu T)-2x_fx_i\bigg),
\end{equation}
which is the classical Euclidean action of the harmonic
oscillator. We now turn to the determinant factor multiplying this
exponential, which is divergent and must be regulated
\emph{before} we can take $m\to0$. Zeta-function regularisation
gives
\begin{equation}
\det\to
  \cosh\bigg(\frac{BT}{m}+2\mu T
  +\mathcal{O}(m)\bigg)-\cosh\bigg(\frac{BT}{m}\bigg).
\end{equation}
up to an $m$ and $T$--independent constant prefactor. The $m\to0$ limit of
the determinant does not exist as there is an essential
singularity at $m=0$. However, as stated in (\ref{AMP21}), we should
subtract the vacuum energy $E_0$ from the Hamiltonian before
taking the limit. This subtraction pre-multiplies the transition
amplitude by $\exp(E_0 T)$. The vacuum energy for this system is
given in \cite{Dunne:1989hv},
\begin{equation}
  E_0 = \sqrt{\frac{B^2}{4m^2} + \frac{\lambda}{m}} =
  \frac{B}{2m} + \mu +\mathcal{O}(m).
\end{equation}
The $m \to 0$ limit of the product of the determinant factor and
$\exp(E_0 T)$ exists (at least for $m \to 0^+$),
\begin{equation}
  \lim_{m \to 0}\,\, \frac{e^{E_0T}}{\det^{1/2}} = e^{\mu
  T/2}\,\,(\sinh(\mu T))^{-1/2}.
\end{equation}
In total we therefore find
\be\fl\eqalign{
  \lim_{m \to 0}&\, \int\! \ud y_f \int\!\ud y_i \, \bra{x_f , y_f} e^{ -
  (\hat{H} - E_0) T} \ket{x_i, y_i} =  \\
&e^{\mu
  T/2}(\sinh(\mu T))^{-1/2}\,\,\exp -\frac{B}{2\sinh(\mu
  T)}\bigg(x_f^2\cosh(\mu T)+x_i^2\cosh(\mu T)-2x_fx_i\bigg),
}
\ee
which is the transition amplitude for the harmonic oscillator,
calculated with the vacuum energy $\mu/2$ subtracted from the
Hamiltonian. This provides a non-trivial example of the
prescription (\ref{AMP21}) describing the transition between a
second and first order theory.

\subsection{A stringy analogue}

The particle in a strong magnetic field has a well known
counterpart system in string theory (see \cite{Ambjorn:2000yr} and
references therein). We consider neutral open strings with
Dp-branes on which the strings end. The worldsheet action is
\be\label{string-act}
  \eqalign{
    S = \frac{1}{4\pi\alpha'}\int\limits_\Sigma\!\ud\sigma& \ud\tau \,
    \,\partial_\tau X^i\, g_{ij}(X)\, \partial_\tau X^j - \partial_\sigma
    X^i\, g_{ij}(X)\, \partial_\sigma X^j \\
    &- \int\!\ud\tau\,\, \partial_\tau X^\mu B_{\mu\nu}
    X^\nu\bigg|_{\sigma=\pi}+\int\!\ud\tau\,\,
    \partial_\tau X^\mu B_{\mu\nu} X^\nu\bigg|_{\sigma=0}.
    }
\ee
Here $g_{\mu\nu}(X)$ and $B_{\mu\nu}(X)$ describe the geometry of
target space. The worldsheet $\Sigma$ for free strings,
parameterised by $\tau$ and $\sigma$, is an infinite strip in the
$\tau$ direction with width $\pi$ in $\sigma$. We will consider
a flat target space, $g_{\mu\nu}(X)=\eta_{\mu\nu}$, and take the two form
flux $B_{\mu\nu}$, representing a magnetic field on the brane, to
be constant.

The equation of motion and boundary conditions in the $p+1$
directions parallel to the brane are
\be
  (\partial_\tau^2-\partial_\sigma^2) X^\mu =0,\quad
  \partial_\sigma X^\mu + 2\pi\alpha' \partial_\tau X^\nu
  {B_\nu}^\mu =0\quad\mathrm{at}\,\,\,\sigma=0,\pi,
\ee
which have the solution \cite{Chu:1998qz}
\be
\fl
  X^\mu=x^\mu +
  p^\mu\tau-2\pi\alpha' p^\nu {B_\nu}^\mu\sigma +
  \sum\limits_{n\not=0}\frac{e^{-in\tau}}{n}\big( i a_n^\mu\cos
  n\sigma - 2\pi\alpha' a_n^\nu {B_\nu}^\mu\sin n\sigma\big).
\ee
The commutation relations are
\be\label{stringCCR}\eqalign{
    \big[x^\mu,x^\nu\big]&= (2\pi\alpha')^2i\,(M^{-1}B)^{\mu\nu}, \\
    \big[x^\mu,p^\nu\big]&= 2\alpha' i\,(M^{-1})^{\mu\nu}, \\
    \big[a_n^\mu,a_m^\nu\big]&= 2\alpha'\,(M^{-1})^{\mu\nu}\,n\,\delta_{n+m},
}
\ee
with all others vanishing and where $M_{ij}=g_{ij}-(2\pi\alpha')^2
B_{ik}g^{kr}B_{rj}$.

There is a low energy limit of this theory which parallels that of
the particle case and which we will discuss shortly. First note
that there is a noncommutativity inherent in this system before we
take any limit. It is straightforward to check using
(\ref{stringCCR}) that
\be
    [X^\mu(\sigma,\tau),X^\nu(\sigma',\tau)]=0,
\ee
for all $\sigma,\sigma'\in(0,\pi)$ and whenever $\sigma$ and
$\sigma'$ are not both $0$ or $\pi$. This means that in the bulk
of target space, away from the branes, spacetime is described by
commutative co-ordinates. However, the commutators between
endpoints of the string are non-zero,
\be\label{endCCR}\eqalign{
    \big[X^\mu(\pi,\tau),X^\nu(\pi,\tau)\big]&=-(2\pi\alpha')^2i \,
    (M^{-1}B)^{\mu\nu}, \\
    \big[X^\mu(0,\tau),X^\nu(0,\tau)\big]&=(2\pi\alpha')^2i \,
    (M^{-1}B)^{\mu\nu}.
}\ee
The ends of the string therefore describe noncommutative
co-ordinates on the branes (closed strings are insensitive to this
effect and see spacetime as a commutative manifold). Let us now
tie this in our to earlier discussions. A low energy limit of this
theory \cite{Seiberg:1999vs} may be taken in which the string
coupling and metric scale as
\be\label{SW}
    \alpha'=\sqrt{\epsilon} \to 0, \qquad g_{\mu\nu} \sim \epsilon \to 0,
\ee
with the magnetic field $B_{ij}$ held fixed --- so this is a limit
in which $B$ is strong compared to other fields, as in the
particle case. In studying this limit it is sufficient to focus on
a pair of co-ordinates so that all metrics and fields become two
by two matrices,
\be
\fl
  g_{\mu\nu}=\epsilon\twomatrix{1 & 0 \\ 0 & 1},\quad
  B_{\mu\nu} = B\twomatrix{0 & 1 \\ -1 & 0},\quad (M^{-1})^{\mu\nu}
  = \frac{1}{\epsilon+4\pi^2B^2}\twomatrix{1 & 0 \\ 0 & 1},
\ee
In this limit, the commutators (\ref{stringCCR})
behave as
\bea
    \big[x^\mu,x^\nu\big] &=& i\frac{4\pi^2 B}{\epsilon+4\pi^2 B^2} \twomatrix{0 & 1 \\
    -1 & 0}\rightarrow -i\big(B^{-1}\big)^{\mu\nu}, \\
    \big[x^\mu,p^\nu\big] &\sim& i \sqrt\epsilon \twomatrix{1 & 0 \\ 0 & 1} ,
    \quad \big[a_n^\mu,a_m^\nu\big] \sim \sqrt\epsilon \twomatrix{1 & 0 \\ 0 & 1},
\eea
so that in this limit the theory undergoes a spectral flow in
which higher energy states of the open string, created by the
action of the oscillators $a^\mu_{-n}$, are suppressed by the
vanishing of $\alpha'$ (which corresponds to the field theory
limit). The scaling of the metric also decouples closed string
states from the theory. We are left with only the lowest energy
states in which the degrees of freedom are the end points of the
open string. In this limit the endpoint commutation relations
(\ref{endCCR}) become
\be \label{endCCR2} \eqalign{
    \big[X^\mu(\pi,\tau),X^\nu(\pi,\tau)\big]&=
    i\twomatrix{0 & -B^{-1} \\ B^{-1} & 0} = i(B^{-1})^{\mu\nu}, \\
    \big[X^\mu(0,\tau),X^\nu(0,\tau)\big] &= -i\twomatrix{0 & -B^{-1} \\
    B^{-1} & 0} = -i(B^{-1})^{\mu\nu},
    }
\ee
in analogy to the quantum mechanical commutator (\ref{B}). Under (\ref{SW}) the first terms of the
action (\ref{string-act}), second order in derivatives, are suppressed and it is only
the boundary terms which survive the limit,
\be
  S \rightarrow -\int\!\ud\tau\,\, \partial_\tau X^\mu B_{\mu\nu}
  X^\nu\bigg|^{\sigma=\pi} + \int\!\ud\tau\,\, \partial_\tau X^\mu B_{\mu\nu}
  X^\nu\bigg|_{\sigma=0},
\ee
again in direct analogy to the particle action (\ref{LLL}). From
this action we may immediately recover (\ref{endCCR2}).

This low energy limit therefore describes a spectral flow in which
higher excitations of the string are suppressed, and from the
action we see that this corresponds to a transition from a second
to first order theory, where the degrees of freedom are
noncommutative particle-like co-ordinates on the branes. It is in
this way that non-commutative field theories arise on the branes
as the low energy limit of string theories \cite{Szabo:2004ic}.

\section{The nonrelativistic limit}\label{NR-sect}

Throughout the remainder of this paper spacetime is $D=1+d$
dimensional unless otherwise stated. Functional integrals over
time dependent fields will be written $\pathD\phi$ while integrals
over configurations at constant time will be written
$\pathD\varphi$.

\subsection{Free theory}

Consider the action of a free relativistic scalar particle given
by the bilinear expression
\be \label{SREL}
  S =\int \! \ud^D x \,\, \phi^\dagger (-\Box - m^2) \phi\,.
\ee
This action describes either a complex scalar field $\phi$ or a
real scalar field $\phi + \phi^\dagger$ (where, for example,
$\phi$ may be taken to be the positive frequency part of the real
field).

In the nonrelativistic limit, all energies and momenta are small
compared to the particle mass $m$. Following \cite{zee:2003} we
define a new field $\Phi$ such that
\begin{equation}\label{newfields}
  \phi(x)=\frac{e^{-imt}}{\sqrt{2m}}\;\Phi(x)\quad \Rightarrow
  \quad\phi^\dagger(x)=\frac{e^{imt}}{\sqrt{2m}}\;\Phi^\dagger(x).
\end{equation}
The motivation for this is that a mode carrying kinetic energy
$E$, oscillates as $\exp -i(E+m)t$. When the kinetic energy is
small compared to $m$ the definitions (\ref{newfields}) factor out
the rapid oscillations which are not admissible in a
nonrelativistic approximation.

This argument can only hold if we restrict the energy and momentum
of the field, so to proceed we work in momentum space. Defining
the Fourier transforms of fields and their conjugates by
\be
  \phi(x)=\int\!\frac{\ud^D p}{(2\pi)^D}\ e^{-i p \cdot x} \,
  \tilde\phi(p),\quad \phi^\dagger(x) = \int\! \frac{\ud^D
  p}{(2\pi)^D}\ e^{i p \cdot x} \, {\tilde\phi}^\dagger(p)\,,
\ee
the action becomes
\be\label{SRELp}
  S = \int\!\frac{\ud^D p}{(2\pi)^D}\,\,{\tilde\phi}^\dagger(p)(p^2-m^2)\tilde\phi(p)\,,
\ee
and one may verify the relations
\be
  \tilde\phi(p)=\frac{1}{\sqrt{2m}}\,\tilde\Phi(p_0-m,\vc{p}),\quad
  {\tilde\phi}^\dagger(p)=\frac{1}{\sqrt{2m}}\,{\tilde\Phi}^\dagger(p_0-m,\vc{p}).
\ee
We now assume that all energies and momenta are small compared to
the particle mass $m$,
\be
  E \equiv p_0 - m \ll m \; , \quad |\vc{p}| \ll m \; ,
\ee
from which we obtain the nonrelativistic approximation
\bea\label{PtoE}
  p^2 - m^2 &=& (E + m - E_p) (E + m + E_p) \nn \\
  &\simeq& (E - \vc{p}^2/2m) (E + 2m + \vc{p}^2/2m) \nn \\
  &\simeq& (E - \vc{p}^2/2m) \, 2m \; .
\eea
We wish to insert this approximation into the action
(\ref{SRELp}). However, it is clear that this is only consistent
if we introduce a large momentum cutoff. Clearly, this is quite
natural in the relativistic theory where we have to regulate
ultra-violet divergences anyhow. Hence, we change variables $E =
p_0 - m$ in (\ref{SRELp}) and impose cutoffs in both $E$ and
momentum $\vc{p}$. For clarity we refer to only a single cutoff
$\Lambda$ with $\Lambda\ll m$. We may now insert our approximation
(\ref{PtoE}) into the action,
\be
  S \to \int^\Lambda
  \frac{\ud E}{2\pi}\frac{\ud^d p}{(2\pi)^d} \,\, \tilde{\Phi}^\dagger(E,\vc{p})\,\bigg(
  E - \frac{\vc{p}^2}{2m}\bigg)\, \Phi(E,\vc{p})\, .
\ee
Inverting the Fourier transform we arrive at the nonrelativistic action in co-ordinate space,
\be
  S\to\int\!\ud^D x\,\ud^D y\,\, \Phi^\dagger(y)\,\delta_\Lambda(x-y)\,\bigg(
  i\partial_t + \frac{\Delta}{2m}\bigg)\, \Phi(x)\, ,
\ee
where short distance divergences are controlled by the regulated delta function
\be\label{delta-reg}
    \delta_\Lambda(x-y) = \int^\Lambda \frac{\ud^Dp}{(2\pi)^D}\,\, \exp -ip \cdot (x - y).
\ee
As we take the cutoff to infinity we arrive at the nonrelativistic
action \cite{Beg:1984yh}
\be \label{SNR}
  S_{\mathrm{NR}} = \int\!\ud^D x\,\,\Phi^\dagger(x)\bigg(i\partial_t+\frac{\Delta}{2m}\bigg)\Phi(x)\; .
\ee
Unlike (\ref{SREL}) this is now linear in the time derivative
$\partial_t \Phi$ and hence the nonrelativistic Schr{\"o}dinger
equation,
\be
  (i\partial_t + \Delta/2m) \Phi(\vc{x},t) = 0 \; ,
\ee
correctly becomes first order in the time derivative. As a result
(and completely analogous to the Dirac field) the Schr{\"o}dinger
matter field does not commute with its conjugate. Rather we read
off the commutator from the $\Phi^\dagger i\partial_t \Phi$ term,
given by $i$ times the Poisson bracket,
\be \label{NRCOMM}
  [\Phi(t,\vc{x}), \Phi^\dagger(t,\vc{y})] = \delta (\vc{x} - \vc{y}) \; .
\ee
Both (\ref{SNR}) and (\ref{NRCOMM}) coincide with the expressions
derived in a slightly different way in the recent text
\cite{zee:2003}.
\subsection{Particle number}
The Schr{\"o}dinger matter field still describes
an (albeit nonrelativistic) many-body theory. However, the
different particle number sectors are separated by huge gaps of
order $m$ so that particle number becomes conserved. Note that
this is also true for nonrelativistic bound states which have
binding energies $E_B$ satisfying
\be
  E_B \ll |\vc{p}|  \ll M \; ,
\ee
where $\vc{p}$ is a typical constituent momentum and $M$ the bound
state mass. Hence, say for a two-particle bound state, we have $M
= 2m - E_B \simeq 2m$, so that $M$ is close to the 2-particle
threshold and hence separated from the one-particle mass-shell by
a gap of almost $m$.

The suppression of number changing interactions in the
nonrelativistic limit may be seen from an action principle.
Reality requires that polynomial interaction terms take one of the
forms
\bea
  S_1 &=\lambda_1\int\!\ud^D x \,\,  \phi^{\dagger n}(x)\phi^n(x)\,,\\
  S_2 &=\lambda_2\int\!\ud^D x \,\,  \phi^{\dagger r}(x)\phi^n(x)+\mathrm{h.c.}\,,\qquad r\not=n.
\eea
Changing to the new fields of (\ref{newfields}), and performing a
change of variables $E=p_0-m$ for each energy integration variable
we have
\be\fl\eqalign{
  S_1 =\lambda_1 \int^\Lambda\tilde{\Phi}^\dagger(E_1)&\ldots
  \tilde{\Phi}^\dagger(E_n)\tilde\Phi(E_{n+1})\ldots \tilde\Phi(E_{2n})\\
  &\times\delta\big(E_1+\ldots +E_n -E_{n+1}-\ldots -E_{2n}\big)\,,
}\ee
\be\fl\eqalign{
  S_2 =\lambda_2
  \int^\Lambda\tilde{\Phi}^\dagger(E_1)&\ldots
  \tilde{\Phi}^\dagger(E_r) \tilde{\Phi} (E_1) \ldots \tilde{\Phi} (E_n) \\
  &\times \delta \big( E_1 + \ldots + E_r -E_1-\ldots -E_n +
  (r-n) m \big) + \mathrm{h.c.} \,, }
\ee
where the integration is over all $E_i$ and $\vc{p}_i$. We have
suppressed the momentum dependence for clarity, and to make a
consistent relativistic approximation we again understand all
integrals to be ultra-violet regulated by $\Lambda$.

In the nonrelativistic approximation, when all energies are small
compared to $m$, we see that the delta function in $S_2$ loses
support because of the non-zero multiple of $m$. These are
precisely the interactions which do not conserve particle number,
or alternatively, which do not conserve nonrelativistic energy.
Hence, only actions of the form $S_1$, which conserve both
particle number and nonrelativistic energy, survive the
nonrelativistic limit.

\subsection{Projection and particle number}

We may now ask, for example, how to get from the field $\Phi$ to
the one-particle sector and the associated Schr{\"o}dinger wave
function $\Psi(x)$? The answer is well known: if $\ket{0}$ and
$\ket{\vc{p}}$ denote the vacuum and a one-particle state of
momentum $\vc{p}$ then, at $t=0$,
\be
  \Psi_{\svc{p}}(x) = \bra{0} \Phi(\vc{x}) \ket{\vc{p}} = e^{i
  \svc{p} \cdot \svc{x}} \;
\ee
is the plane wave solution of the single-particle Schr\"odinger
equation for a free nonrelativistic particle.

To proceed on a slightly more formal level we expand the
relativistic field operator (specialising to $D=2$) in a Fock basis,
\be\label{field}
  \phi (t,x) = \int\!\frac{\ud k}{2\pi} \frac{1}{\sqrt{2\,
  \omega_k}} \left[ a_k e^{-i \omega_k t+i k x} + a^\dagger_k e^{i
  \omega_k t-i k x}.  \right] \; ,
\ee
with $w_k \equiv \sqrt{k^2+m^2}$ and the Fourier modes obeying
$[a_q,a^\dagger_k]=2\pi\delta(k-q)$. The field operator $\phi$
changes particle number by one unit, and a one particle state
$\ket{k}$ with momentum $k$ is defined by
\be
  \ket{k} =  \sqrt{2\omega_k}\  a^\dagger_k\ket{0} \quad
  \Rightarrow \quad \bracket{q}{k} = 4\pi\,\omega_k\,\delta(q-k).
\ee
Hence, it makes sense to consider the truncated fields
\bea
  \phi_{01} (t,x) &\equiv& \PP_0 \phi(t,x) \PP_1 \; , \label{PHI01} \\
  \phi_{10} (t,x) &\equiv& \PP_1 \phi(t,x) \PP_0 \; , \label{PHI10}
\eea
where we have introduced the projections onto vacuum and
one-particle sectors, respectively,
\be
  \PP_0 \equiv \ket{0} \bra{0},\qquad \PP_1  \equiv
  \int\limits_{-\Lambda}^\Lambda\!\frac{\ud k}{2\pi}\,\,
  \ket{k}\frac{1}{2\omega_k}\bra{k}\; ,
\ee
and we restrict the range of momentum to $|k|\leq\Lambda\ll m$ in
accordance with the nonrelativistic approximation. We then infer
the $t=0$ commutator of the projected fields
\be
  [\phi_{01}(t,x), \phi_{10}(t,y)] = \PP_0 \phi(t,x) \PP_1 \phi(t,y) \PP_0
  - \PP_1 \phi(t,y) \PP_0 \phi(t,x) \PP_1 \; ,
\ee
with the first (second) term obviously acting in the vacuum (one-particle)
sector. Projecting onto the former we find
\bea
  \PP_0 [\phi_{01}(t,x), \phi_{10}(t,y)] \PP_0 &=& \int\limits_{-\Lambda}^\Lambda\!
  \frac{\ud k}{2\pi}\ \frac{1}{2\omega_k}\ e^{ik(x-y)}\PP_0 \\
  &=& \frac{1}{2m}\int\limits_{-\Lambda}^\Lambda\!\frac{\ud k}{2\pi}\
  \frac{1}{\sqrt{1+k^2/m^2}}\ e^{ik(x-y)}\PP_0.
\eea
For $|k|\leq\Lambda\ll m$ the square root is approximately unity,
and rescaling the fields with $\sqrt{2m}$ we find the following
analogue of (\ref{PXYP}),
\be
  2m \, \PP_0 [\phi_{01}(t,x), \phi_{10}(t,y)] \PP_0 \simeq
  \int\limits_{-\Lambda}^\Lambda\!\frac{\ud k}{2\pi}\
  e^{ik(x-y)} \, \PP_0 = \frac{\sin \Lambda(x-y)}{\pi(x-y)} \, \PP_0.
\ee
The right-hand side is the one dimensional regulated delta
function (\ref{delta-reg}) multiplying the projection $\PP_0$.
Including the rescaling we thus identify
$\sqrt{2m}\,\phi_{01}\sim\Phi$ and
$\sqrt{2m}\,\phi_{10}\sim\Phi^\dagger$, recovering (\ref{NRCOMM})
in the vacuum sector of the nonrelativistic theory.

Note that the projection formalism above is quite reminiscent of
the old Tamm-Dancoff idea of truncating in particle number
\cite{Tamm:1945qv,Dancoff:1950ud}. If we expand the projected
fields (\ref{PHI01}) and (\ref{PHI10}) we obtain explicitly
\bea
  \phi_{01} (t,x) &=& \int\limits_{-\Lambda}^\Lambda\!\frac{\ud k}{2\pi}\
  \frac{1}{2\omega_k} e^{-i\omega_k t+ik x} \ket{0} \bra{k} \; , \\
  \phi_{10} (t,x) &=& \int\limits_{-\Lambda}^\Lambda\!\frac{\ud k}{2\pi}\
  \frac{1}{2\omega_k} e^{i\omega_k t-ik x} \ket{k} \bra{0} \; .
\eea
which corresponds to a cutoff in particle number, $N \le 1$. In
other words, one essentially projects onto negative (positive)
frequencies or the annihilation (creation) parts of the field,
replacing
\be
  a_k \to \ket{0} \bra{k} \; , \quad a_k^\dagger \to \ket{k}
  \bra{0} \; ,
\ee
in the Tamm-Dancoff spirit. However, it seems obvious that this
can only be consistent in a nonrelativistic context where energies
and momenta are small compared to particle masses; in a
relativistically covariant theory any large boost will spoil these
scale hierarchies as boosts, being dynamical Poncar{\`e}
generators \cite{Dirac:1949cp}, neither conserve energy nor
particle number. We will briefly come back to these issues in the
next section.

\subsection{Transition amplitudes}

What can we say about the behaviour of quantum amplitudes in the
relativistic limit? Such amplitudes are described in the
Schr\"odinger picture by gluing state wave functionals onto the
Schr\"odinger functional $\bra{\varphi_f} e^{-i\hat{H}T}
\ket{\varphi_i}$ which generates time evolution (see appendices A
and B),
\be
  \bra{\Psi_2} e^{-i \hat{H}T} \ket{\Psi_1} = \int
  \pathD\varphi_f\mathscr{D}\varphi_i \, \, \Psi^*_2 [\varphi_f]
  \bra{\varphi_f}e^{-i\hat{H}T}\ket{\varphi_i} \Psi_1[\varphi_i].
\ee
As described in the appendices the Schr\"odinger functional is
characterised by temporal boundary terms which, in second order
theories, depend on both the field and its derivative, reflecting
the fact that Cauchy data is required to determine time evolution.
In first order theories the boundary terms depend on the field and
its conjugate and do not contain time derivatives (which are not
required as data). We have seen how the action changes from a
second to first order theory in the nonrelativistic limit, so let
us now turn to these boundary terms.

A typical relativistic boundary term for a real scalar field,
imposing the Dirichlet condition $\phi(T,\vc{x}) =
\varphi(\vc{x})$ at time $t=T$ is
\be\label{lump}
  i\int\!\ud^d x \,\, \varphi(x)\dot{\phi}(T,x) +
  \frac{i}{2}\int\!\ud^d x\,\,\Lambda'\varphi^2(x),
\ee
where $\phi(t,\vc{x})$ is a functional integration variable
obeying the boundary condition $\phi(T, \vc{x})=0$,
$\varphi(\vc{x})$ is the boundary field and $\Lambda'$ is a
regularisation of $\delta(0)$. As first noted by Stueckelberg
\cite{Stueckelberg:1951} (see also \cite{Bogolubov:1959}) and
discussed in detail by Symanzik, \cite{Symanzik:1981wd}, placing
sources on the boundary leads to divergences in perturbation
theory when the field and its `image charges' (which impose the
boundary conditions on propagators) are placed at the same point
in time --- $\Lambda'$ is hence a short time regulator, which may
also be seen as a UV regulator of the field momentum propagator
(see appendix B).

In terms of the nonrelativistic degrees of freedom the terms
(\ref{lump}) may be written, including a rescaling of the boundary
data, $\varphi\rightarrow\varphi/\sqrt{2m}$,
\be\label{terms}\eqalign{
    \int\!\ud^d x\,\,&\varphi(\x) \, \big[ e^{-imT}\Phi(T,\x)-e^{imT}
    \Phi^\dagger(T,\x)\big] \\
    &+ \frac{i}{m}\int\!\ud^d x\,\,\varphi(\x) \, \big[ e^{-imT}
    \partial_t\Phi(T,\x)+e^{imT} \partial_t\Phi(T,\x)^\dagger \big] \\
    &+ \frac{i}{2m}\int\!\ud^d x\,\,\Lambda'\varphi^2(\x).
}\ee
Transforming to momentum space and imposing a cutoff, we see that
the second line of (\ref{terms}) goes like $E/m$ and is hence
suppressed in the nonrelativistic limit. As discussed above, this
is consistent with expectations because boundary terms in first
order theories should not depend on derivatives of the fields. The
third line of (\ref{terms}), in momentum space, goes like
$\Lambda'/m$. Such terms are also absent in first order theories
and we see that if we maintain $m$ as the largest scale in our
theory, this term is also suppressed in the nonrelativistic limit.

We are therefore left with the first line of (\ref{terms}),
depending on the fields and not their momenta, with the condition
\be
    \phi(T,\x) = 0\rightarrow e^{-imT}\Phi(T,\x)+e^{imT}\Phi^\dagger(T,\x)=0,
\ee
up to small corrections, implying a mixed boundary condition on
the nonrelativistic fields. We will discuss further limits of
field theory wave functionals towards the end of the next section.

\section{Light-front quantisation}\label{LF-sect}

In 1949 Dirac pointed out that, in a relativistically
covariant theory, there are several alternative ``forms of
relativistic dynamics'' \cite{Dirac:1949cp}. In particular, one
may postulate field commutators on null planes rather than
entirely space-like hypersurfaces leading to light-cone or,
somewhat more precisely, light-front quantum field theory. The
literature on this subject is vast and we refer the reader to the
reviews
\cite{Namyslowski:1985zq,Burkardt:1995ct,Brodsky:1997de,Heinzl:2000ht}
and the references cited therein.

One of the hopes of studying light-front quantum field theory was
indeed that the Tamm-Dancoff approximation of the previous section
might become feasible \cite{Perry:1990mz,Perry:1991ny} in a
relativistic context. This hope was based on the fact that, upon
quantising on null-planes, a number of nonrelativistic features
seem to arise within a fully relativistic approach. This was first
noted by Weinberg in his analysis of the infinite momentum limit
of Feynman graphs \cite{Weinberg:1966jm} (see also
\cite{Heinzl:2006dw}) and can be made explicit in terms of a 2$d$
Galilei subgroup of the Poincar{\'e} group
\cite{Susskind:1967rg,Bardakci:1969dv}. Among the consequences one
finds, for example, a separation of relative and centre-of-mass motion
within bound states. Most interesting seems to be the closely
related suppression of vacuum fluctuation and pair production
effects expressed as the folkloric statement that the
light-front vacuum is `trivial'
\cite{Burkardt:1995ct,Brodsky:1997de,Heinzl:2000ht}. In what
follows we will take a fresh look at the nonrelativistic
aspects of light-front field theory in terms of spectral flow.

\subsection{Time-slice geometry}

Field quantisation on an arbitrary hypersurface (with time-like or
light-like normals) may be formulated as follows
\cite{Heinzl:2000ht}. We introduce a coordinate transformation $x
\to \xi$ (and likewise for momenta, $p \to k$),
\be \label{COORDTRAFO}
  \xi^\alpha = L^\alpha_{\;\;\mu} \, x^\mu \; , \quad k_\alpha =
  L_\alpha^{\;\;\mu} \, p_\mu \; , \quad L^\alpha_{\;\;\mu}
  L_\beta^{\;\;\mu} = \delta^\alpha_{\;\;\beta} \; .
\ee
The new variables describe an alternative (3+1)-foliation of
Minkowski space with $\xi^0$ being the new time variable,
conjugate to the momentum component $k_0$. We assume that the
transformation is linear\footnote{Clearly, this is not the most
general case. Even in special relativity one can choose
hyperboloids rather than planes as surfaces of equal time which
corresponds to Dirac's `point-form' of relativistic dynamics
\cite{Dirac:1949cp}.} so that
\be
  \Sigma: \xi^0 = \mathrm{const}
\ee
is a hyperplane of equal time $\xi^0$. The metric
associated with the transformation (\ref{COORDTRAFO}) is
\be \label{METRIC31}
  g_{\alpha\beta} = \eta_{\mu\nu} \, L^\mu_{\;\; \alpha} \,
  L^\nu_{\;\; \beta} \equiv
                    \left( \begin{array}{cr}
                            g_{00}  & \vc{g}^T \\
                            \vc{g} & -G
                      \end{array}        \right) \; ,
\ee
where we have introduced a (3+1)-split in the last step. Hence,
$-G$ is the induced metric on the quantisation hyperplane. The
(constant) normal on $\Sigma$ is
\be
  N_\mu = \left. \pad{\xi^0}{x^\mu} \right|_\Sigma = L^0_{\;\;\mu} \; ,
\ee
and prominently enters the inverse metric which we write as
follows,
\be
   \label{INVMETRIC31}
   g^{\alpha\beta} = \left( \begin{array}{cr}
                            N^2     & \vc{\gamma}^T \\
                            \vc{\gamma} &  -\Gamma
                      \end{array}        \right)\; .
\ee
The square of the normal can be expressed in terms of metric
determinants from (\ref{METRIC31}),
\be
  N^2 = g^{00} = \det (-G)/ \det(g) \; ,
\ee
and will become important in a moment. The inverse metric
(\ref{INVMETRIC31}) governs the mass-shell constraint,
\be \label{CONSTRAINT}
  0 = p^2 - m^2 = g^{\alpha\beta} k_\alpha  k_\beta - m^2 \; ,
\ee
which will be used to determine the energy variable $k_0$ in terms
of the `spatial' components $k_a$, $a = 1,2,3$. We mention in
passing that introducing the space-time foliation $(\xi^0, \xi^a)$
can be viewed as gauge-fixing the time reparametrisation
invariance $\tau \to \tau'$ generated by the constraint
(\ref{CONSTRAINT}). The associated Faddeev-Popov (FP) expression
is \cite{Heinzl:2000ht}
\be \label{FP}
  \FP \equiv N \cdot p \equiv N^2 k_0 + \gamma^a k_a \; ,
\ee
which has to be evaluated on mass-shell, i.e.\ by expressing $p_0$
and $k_0$ in terms of $p_i$ and $k_a$, respectively via
(\ref{CONSTRAINT}). Expanding the latter we find the quadric
\be \label{QUADRIC}
  0 = N^2 k_0^2 + 2 \gamma^a k_a k_0 - \Gamma^{ab} k_a
  k_b - m^2 = 0 \; .
\ee
Interestingly, its discriminant $\Delta$ basically coincides with
the FP expression squared,
\be \label{FPDELTA}
  \Delta \equiv 4 \, \Big\{ (\gamma^a k_a)^2 + N^2 (\Gamma^{ab}
  k_a k_b + m^2) \Big\} = 4 \, \FP^2 \; .
\ee
Depending on the value of $N^2$, we thus have to consider two
different cases. The generic one is that the normal $N$ on
$\Sigma$ is \textit{time--like}, $N^2 > 0$. In this case, the
mass--shell constraint is of \textit{second order} in $k_0$, so
that there are two distinct solutions,
\be \label{2NDORDER}
  k_0 = \frac{1}{N^2} \, \Big( - \gamma^a k_a \pm
  \sqrt{\Delta}/2 \Big) =  \frac{1}{N^2} \, \Big( - \gamma^a k_a \pm
  |\FP| \Big) \; .
\ee
The second case to be considered is in a sense degenerate. It
corresponds to a \textit{light--like} normal, $N^2 = 0$. In this
case, the constraint (\ref{QUADRIC}) is only of \textit{first
order} in $k_0$ leading to a single solution,
\be \label{1STORDER}
  k_0 = \frac{\Gamma^{ab} k_a k_b + m^2}{2 \, |\FP|} \; .
\ee
with $\FP = \gamma^a k_a$ according to (\ref{FP}). While
(\ref{1STORDER}) looks simpler than (\ref{2NDORDER}) (unique sign,
no square root) a new difficulty arises as the momentum projection
$\FP = \gamma^a k_a$ may be vanishing. We thus have a Gribov
problem which will turn into the notorious zero mode problem of
light-front field theory.

In what follows we want to study the light-like limit (LLL), $N^2
\to 0$.

\subsection{The LLL metric}

To the best of our knowledge a limiting approach to light-front
coordinates was first suggested by Chen in 1971
\cite{Chen:1971yg}. His new coordinates differed  from light-front
ones by an infinitesimal rotation. Finite rotations were later
considered in \cite{Frishman:1976vi,Hornbostel:1991qj}.

Clearly, rotations preserve the orthogonality of the coordinates.
For relativistic systems this is not a crucial issue, however, and
one may as well give up orthogonality. This approach was first
adopted by Prokhvatilov and Franke \cite{Prokhvatilov:1989eq} and
independently by Lenz et al.\ \cite{Lenz:1991sa}. Since then it
has frequently been utilised for field theory applications both at
zero and finite temperature (see e.g.\
\cite{Prokhvatilov:1994dm,Naus:1997zg,Ilgenfritz:2006ir} and
\cite{Alves:2002tx,Das:2004je}, respectively). It has also been
adopted for the matrix model approach to M-theory
\cite{Hellerman:1997yu} where the notion of the `LLL' was coined.
The idea is to introduce the new coordinates
\be \label{PFTRANS}
\eqalign{
  \xi^0 &= x^0 \left(1 + \eta^2/2 \right) + x^3 \left(1 - \eta^2/2
  \right) \; , \\
  \xi^3 &= x^\m \equiv x^0 - x^3 \; , \\
  \xi^{1,2} &= x^{1,2} \; .
} \ee
such that, in the limit $\eta \to 0$, $\xi^0$  becomes the
standard light-front time,
\be
  x^\p = x^0 + x^3 \; .
\ee
The invariant distance element is
\be
  ds^2 = d\xi^0 d\xi^3 - d\xi_\perp^2 - \frac{\eta^2}{2} d\xi^3
  d\xi^3 \; ,
\ee
implying the following metric and its inverse
\be \label{PFMETRIC} \fl
  g_{\alpha \beta} = \left( \begin{array}{cccc}
                            0 & 0 & 0 & 1/2 \\
                            0 & -1 & 0 & 0 \\
                            0 & 0 & -1 & 0 \\
                            1/2 & 0 & 0 & - \eta^2/2
                            \end{array}
                            \right) \; , \quad
  g^{\alpha\beta} = \left( \begin{array}{cccc}
                            2\eta^2 & 0 & 0 & 2 \\
                            0 & -1 & 0 & 0 \\
                            0 & 0 & -1 & 0 \\
                            2 & 0 & 0 & 0
                            \end{array}
                            \right) \; ,
\ee
which we will henceforth refer to as the LLL metric. Comparing
with (\ref{INVMETRIC31}) we read off that the hyperplane normal
$N$ on $\Sigma: \xi^0 = 0$ satisfies
\be
  N^2 = 2 \eta^2 > 0 \;,
\ee
implying that $N$ is time-like for $\eta \ne 0$. In accordance
with that, the line element on $\Sigma$ is space-like even for
vanishing transverse separation,
\be
  \left. ds^2\right|_{\xi^0 = 0 = \xi_\perp} = - \frac{\eta^2}{2} \, d\xi^3
  d\xi^3 < 0 \; .
\ee
Hence, for $\eta \ne 0$, $\Sigma$ is indeed a space-like
hyperplane.
\begin{figure}[!ht]
\centering\includegraphics[angle=-90,scale=0.6]{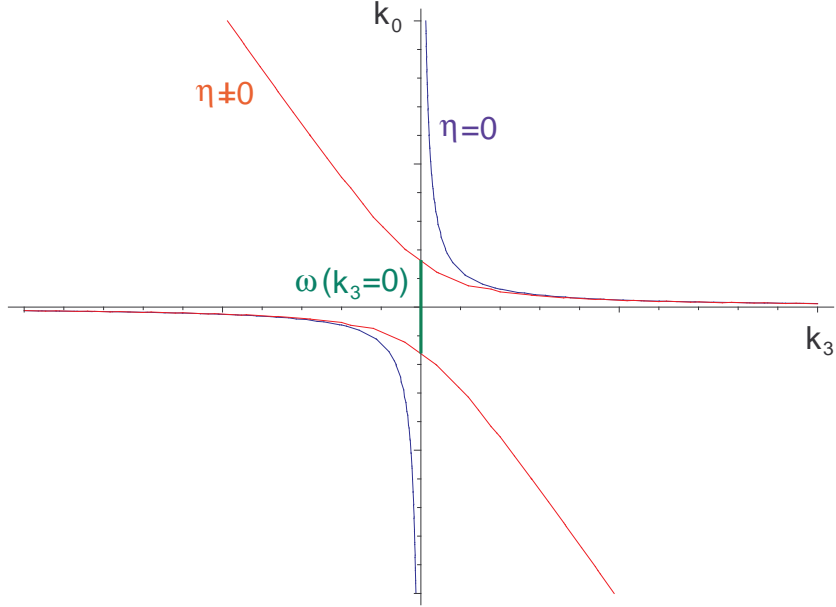}
\caption{Mass shells for $\eta \ne 0$ and in the LLL ($\eta = 0$)
where the signs of $k_0 = p_+$ and $k_3 = p_-$ coincide. Also
displayed is the energy gap $\omega(k_3 = 0)$.} \label{ShellFig}
\end{figure}
Figure \ref{ShellFig} shows the mass shell energy-momentum
relation (\ref{CONSTRAINT}) in our co-ordinates (\ref{PFTRANS})
and in light-front co-ordinates. The on-shell energies at
$\eta\not=0$ are
\be\label{EonShell}
    \omega^\pm = -\frac{k_3}{\eta^2}\pm\sqrt{\frac{k_3^2}{\eta^4}
    + \frac{k_\perp^2+m^2}{2\eta^2}}.
\ee
The difference between these two energies will be denoted
$\omega$,
\be
    \omega:=\omega^+-\omega^-=2\sqrt{\frac{k_3^2}{\eta^4}+\frac{k_\perp^2+m^2}{2\eta^2}}\,,
\ee
where $\omega(k_3=0)$ is the size of the gap in the energy
spectrum of figure \ref{ShellFig}. Expanding the on-shell
energies,
\be\label{cases}
  \omega^\pm = \left\{\eqalign{ \frac{ k_\perp^2+m^2}{4 k_3}
  + \mathcal{O}(\eta^2)\quad\mathrm{if}\quad k_3\gtrless 0 \\
  \pm\frac{1}{\eta}\sqrt{\frac{ k_\perp^2+m^2}{2}}\quad\mathrm{if}\quad k_3=0 \\
  \frac{-2 k_3}{\eta^2}-\frac{ k_\perp^2+m^2}{4 k_3} + \mathcal{O} (\eta^2)
  \quad \mathrm{if}\quad k_3\lessgtr 0}\right.
\ee
and referring to the four quadrants of figure \ref{ShellFig}
enumerated anticlockwise from the upper right, we see that the
energies in the first and third quadrant remain finite as
$\eta\to0$ and become the expected light-front energies, while
those in the second and fourth quadrants flow to infinity. In the
following subsection we will demonstrate explicitly how this
spectral flow gives rise to noncommutativity in the fields.

\subsection{Poisson Brackets of the field}\label{PoisSect}

A scalar field obeying the LLL co-ordinate equations of motion may
be written
\be\fl\eqalign{
  \phi(\xi^0,\xi^a)&=\int\!\frac{\ud k_0\ud^3k}{(2\pi)^4}\
  e^{-ik_\mu\xi^\mu}\tilde\phi(k_0,k_a)\,2\pi\delta \bigg[
  \eta^2k_0^2+2k_0k_3-\frac{1}{2}(k_\perp^2+m^2)\bigg] \\
  &=\int\!\frac{\ud^3k}{(2\pi)^3}\ e^{-i\omega^+\xi^0-ik_a\xi^a}
  \frac{\tilde\phi(\omega^+,k_a)}{\eta^2\omega} +
  e^{-i\omega^-\xi^0-ik_a\xi^a}\frac{\tilde\phi(\omega^-,k_a)}{\eta^2\omega}.
  }
\ee
Here $\omega^\pm$ are the on shell energies (\ref{EonShell}). The
conjugate field momentum is given by
\be\eqalign{
    \Pi &= \eta^2\partial_0\phi+\partial_3\phi \\
     &=\int\!\frac{\ud^3 k}{(2 \pi)^3}\,\, - \frac{i}{2}
     e^{-i\omega^+\xi^0-i k_a\xi^a}\tilde\phi(\omega^+, k_a) +
     \frac{i}{2}e^{-i\omega^-\xi^0-i k_a\xi^a}\tilde\phi(\omega^-, k_a).
     }
\ee
where the first line shows that the momentum contains the velocity
$\partial_0 \phi$ only as long as $\eta \ne 0$. For $\eta = 0$,
however, $\Pi$ is merely an abbreviation for the spatial
derivative, $\partial_3\phi$. It is easily verified that the equal
time Poisson brackets
\bea
    \big[\phi(\xi^0,\xi^a),\Pi(\xi^0,{\xi'}^a)\big]&=&i\delta^3(\xi^a-{\xi'}^a)\,, \\
    \label{fieldfield}\big[\phi(\xi^0,\xi^a),\phi(\xi^0,{\xi'}^a)\big]&=&0\,,
\eea
are equivalent to
\be
  \big[\tilde\phi(\omega^+, k_a),\tilde\phi(\omega^-,
  k'_a)\big]=(2 \pi)^3\,\eta^2\omega\,\delta^3( k_a+ k'_a), \ee
with all other brackets vanishing. Referring again to the four
quadrants of figure \ref{ShellFig}, we see that the bracket is
non-zero only when one momentum space field has support in the
first (second) quadrant and one in the third (fourth). It is
precisely the cancellation between these two pairs of sectors
which makes the Poisson bracket (\ref{fieldfield}) of the field
with itself vanish. This may be verified by splitting the field
into two parts, defined by the range of integration over $ k_3$,
as so,
\be\fl\eqalign{
  \phi(\xi^0,\xi^a)&=\int\!\frac{\ud^2 k_\perp}{(2\pi)^2}
  \int\limits_0^\infty\!\frac{\ud k_3}{2\pi}\,\,
  e^{-i\omega^+\xi^0-i k_a\xi^a}\frac{\tilde\phi(\omega^+, k_a)}{\eta^2\omega}+
  \int\limits_{-\infty}^0\!\frac{\ud k_3}{2\pi} \,\, e^{-i\omega^ - \xi^0-i k_a\xi^a}
  \frac{\tilde\phi(\omega^-, k_a)}{\eta^2\omega} \\
  &+\int\!\frac{\ud^2 k_\perp}{(2\pi)^2} \int\limits^0_{-\infty}
  \!\frac{\ud k_3}{2\pi}\,\, e^{-i\omega^+\xi^0-i k_a\xi^a}
  \frac{\tilde\phi(\omega^+, k_a)}{\eta^2\omega}+ \int\limits^\infty_0\!
  \frac{\ud k_3}{2\pi}\,\, e^{-i\omega^-\xi^0-i k_a\xi^a}
  \frac{\tilde\phi(\omega^-, k_a)}{\eta^2\omega}.
    }
\ee
These terms live in quadrants one, three, two and four
respectively. The contribution from quadrants one and three to the
commutator $[\phi,\phi]$ is
\be\label{stays}
  \int\!\frac{\ud^3 k_a}{(2\pi)^3}\,\mathrm{Sign}( k_3)\,\,e^{-i
  k_3(\xi-\xi')^3-i k_\perp.(\xi-\xi')^\perp}\frac{1}{\eta^2\omega},
\ee
while quadrants two and four contribute \be -\int\!\frac{\ud^3
k_a}{(2\pi)^3}\,\mathrm{Sign}( k_3)\,\,e^{-i k_3(\xi-\xi')^3-i
k_\perp.(\xi-\xi')^\perp}\frac{1}{\eta^2\omega}, \ee which
together cancel to give (\ref{fieldfield}). Thus, the
commutativity of the fields on hypersurfaces of equal time $\xi^0$
(expressing relativistic causality) is actually resulting from a
delicate `interplay' of different field modes. We will see in a
moment that this `interplay' depends crucially on the flow
parameter $\eta$.

So let us now take the limit of vanishing $\eta$. As $\eta \to 0$
terms in $\phi$ corresponding to quadrants one and three have well
defined limits, as may be read off from (\ref{cases}). The other
terms, however, contain rapidly oscillating complex exponentials
since $\omega^\pm\sim k_3\eta^{-2}$ in these quadrants (and
$\omega^\pm\sim\eta^{-1}$ on the boundary $k_3=0$) and will be
suppressed by said oscillations. Observing that $\eta^2\omega$ is
finite as $\eta\to0$, $\eta^2 \omega \to 2 |p_-|$, we are left
with a truncated field,
\be\fl
  \phi(x^+,x^a)=\int\!\frac{\ud^2p_\perp}{(2\pi)^2}\int\limits_0^\infty\!\frac{\ud
  p_-}{2\pi}\,\, e^{-iEx^+ -i
  p_ax^a}\frac{\tilde\phi(E,p_a)}{2|p_-|}+\int\limits_{-\infty}^0\!\frac{\ud
  p_-}{2\pi}\,\, e^{-iE
  x^+-ip_ax^a}\frac{\tilde\phi(E,p_a)}{2|p_-|},
\ee
where we have written
\be
    E\equiv E(p_-,p_\perp)=\frac{p_\perp^2+m^2}{4p_-} = \frac{p_\perp^2+m^2}{2p^+}.
\ee
The momentum space commutator is also well defined in this limit
for $p_->0$, $p'_-<0$,
\be\label{newCCR}
    \big[\tilde\phi(E,p_a),\tilde\phi(E,p'_a)\big] =
    2|p_-|\,(2\pi)^3\, \delta^3(p+p'),
\ee
and recalculating the Poisson bracket of the field with itself
(this may also be read off from the $\eta\to0$ limit of
(\ref{stays})) we find
\be\fl\eqalign{
    \big[\phi(x^+,x^a),\phi(x^+,{x'}^a)\big] &= \int\!\frac{\ud^2p_\perp
    \ud p_-}{(2\pi)^3}\ \mathrm{Sign}(p_-)\,\,e^{-ip_a(x-x')^a}\frac{1}{2|p_-|} \\
    &=\delta^\perp(x-x')\int\limits_{-\infty}^\infty\!
    \frac{\ud p_-}{2\pi}\,\,\frac{1}{2p_-}\ e^{-ip_-(x^--{x'}^-)} \\
    &=-\frac{i}{4}\delta^\perp(x-x')\ \mathrm{Sign}(x^--x{'^-}).
    }
\ee
Thus, we have finally arrived at the canonical Poisson bracket of
the light-front field. Again, we see explicitly that spectral
flow, causing the decoupling of high-energy states from the
theory, alters the Poisson brackets, and therefore the
commutators, of the theory. Following the flow all the way to the
LLL one goes from a second to a first order theory, thereby
inducing a noncommutativity in the configuration space of the
original system.

It is useful to work in a mixed representation of light-front
theory where the field depends on $x^+, p_-$ and $x^\perp$ by
defining, for $p\equiv p_->0$,
\be\eqalign{
     \phi_p(x^+,x^\perp) &\equiv \sqrt{2|p|}\int\!\frac{\ud^2 p_\perp}{(2\pi)^2}\
     e^{-iE x^+-ip_\perp x^\perp}\tilde\phi(E,p_a)\,, \\
    \phi^\dagger_{p}(x^+,x^\perp) &\equiv \phi_{-p}(x^+,x^\perp)\,.
}
\ee
Using the commutation relations (\ref{newCCR}) it is easily
checked that
\be\label{lc-comm}
    \big[\phi_p(x^+,x^\perp),\phi^\dagger_q(x^+,{x'}^\perp)\big]
    = \delta^\perp(x-x')\,\delta(p-q).
\ee
So, for all $x^ + \ne 0$ the limit $\eta \to 0$ takes us, via
spectral flow, to the light-front theory. Note though that our
manipulations only hold for $\xi^0 \ne 0$, that is \textit{off}
the quantisation hyperplane. Degrees of freedom are eliminated by
rapid oscillations only for $\xi^0 > 0$ but remnants survive at
$\xi^0 = 0$ $(x^+ = 0)$ in the limit. There are therefore extra
degrees of freedom which remain in the quantisation surface $x^+ =
0$ and do not propagate into the bulk, $x^+ > 0$. It seems
plausible that the boundary degrees of freedom are related to the
notorious light-front zero modes (reviewed in
\cite{Yamawaki:1998cy}) as their propagator is instantaneous,
namely proportional to $\delta (x^+)$ \cite{Heinzl:2003jy} and
hence indeed located at the temporal boundary. In the following
subsections we will see that the same distinction between bulk and
boundary arises in the functional picture.

\subsection{Light-front limit of wave functionals - the vacuum}

As in previous sections additional insight is provided by studying
the behaviour of wave functionals in the light-front limit. We
begin with the $\eta^2 \not= 0$ vacuum wave functional
$\Psi_0[\varphi]$, which may be written as a sum over all field
histories beginning in the infinite past and intersecting the
configuration $\varphi(x)$ at time $\xi^0=0$ (see
\cite{Jaramillo:1998ff}, \cite{Birmingham:1986gj},
\cite{Hartle:1983ai} for applications in field theory, string
theory and quantum gravity),
\be\label{vaccy}
  \Psi_0[\varphi]=\int\pathD\phi\,\,\exp
  -\int\limits_{-\infty}^0\ud\xi^0\,\,L[\phi]\bigg|^{\phi=\varphi\
  \mathrm{at}\ \xi^0=0}\,,
\ee
where we have rotated to Euclidean space, $\xi^0 \to -i\xi^0$, and
$L[\phi]$ is the (free) Euclidean Lagrangian. The boundary
condition at $\xi^0=-\infty$ is that the field should be regular.

The integral is computed by splitting $\phi$ into a classical part
which obeys the equation of motion and boundary conditions, and a
quantum fluctuation which obeys a Dirichlet boundary condition at
$\xi^0=0$. The general solution of the equations of motion is
\be\label{cl-field}
  \phi = \int\!\frac{\ud^3 k}{(2\pi)^3}\,\,e^{-ik_a \xi^a}
  e^{-\xi^0\omega^+}\tilde\phi_1(k_a) + e^{-ik_a \xi^a}
  e^{-\xi^0\omega^-}\tilde\phi_2(k_a),
\ee
where $\omega^\pm$ are the on-shell energies of (\ref{EonShell}).
The boundary conditions of (\ref{vaccy}) imply that
$\tilde\phi_1=0$ and $\tilde\phi_2=\tilde\varphi$. The classical
and quantum pieces are `orthogonal' in that the action splits into
two copies, one evaluated with the above solution and one
evaluated with the quantum fields. In this way the integral may be
performed to find
\be
  \Psi_0[\varphi] = Z(\eta)\exp
  \frac{1}{2}\int\! \ud^3 \xi \, \ud^3 \xi' \, \varphi(\xi^a)
  \, \Omega(\xi^a,{\xi'}^a) \, \varphi({\xi'}^a)
\ee
with covariance
\be\label{Omega-def}
  \Omega(\xi^a,{\xi'}^a) = \int\! \frac{\ud^3
  k}{(2\pi)^3} \, \, e^{-ik_a
  (\xi^a - {\xi'}^a)} \, \frac{\eta^2}{2} \, \big[ \,\omega^-(k)
  - \omega^+ (k) \big],
\ee
and where $Z(\eta)$ is the contribution of quantum fluctuations.
This may be determined from the normalisation condition
$|\Psi_0|^2 = 1$, which implies
\be \fl
  Z^2(\eta) = \frac{1}{2}\int\pathD\varphi\,\,\exp -
  \frac{1}{2}\int\! \ud^3 \xi \, \ud^3 \xi' \, \varphi(\xi^a) \, [
  \Omega(\xi^a,{\xi'}^a)+\Omega^\dagger(\xi^a,{\xi'}^a)] \, \varphi({\xi'}^a).
\ee
In the limit $\eta \to 0$, the product $\eta^2\omega^-(k_a)$
vanishes for all $k_3 \geq 0$ and tends to $-2k_3$ for $k_3<0$. We
therefore find the LLL vacuum wave functional $\Psi_0$,
\be \fl \eqalign{
  Z(0)\exp -\int\! \ud^3 x \, \ud^3 x'\,\, \varphi(x) \varphi(x')
  \int\!\frac{\ud^2 p_\perp}{(2\pi)^2}\int\limits_0^\infty\!
  \frac{\ud p_-}{2\pi}\,\,e^{-ip_\perp (x^\perp-{x'}^\perp)}
  p_- e^{-ip_-(x^--{x'}^-)} \; , \nn \\
  }
\ee
which may be simplified to give
\be \label{LLLVAC}
  \Psi_0 [\varphi] = \mathrm{Det}^{1/4}(i\partial_3) \, \exp - \int\limits_0^\infty
  \! \frac{\ud p}{2\pi} \int \! \ud^2 x^\perp \,\, \varphi_{-p}(x^\perp)
  \,p \, \varphi_p (x^\perp)  \; .
\ee
Here we have again transformed to the mixed representation with
$p\equiv p_-$ and $\varphi_{-p} = \varphi_p^\dagger $. The result
(\ref{LLLVAC}) coincides with the light-front vacuum wave
functional for free scalar fields found in \cite{Heinzl:2000ht}.
Some remarks are in order at this point. First, one notes the
interesting property that the covariance is local,
\be \label{LLLCOV}
  \Omega (p, \, p'; x^\perp, x^{\prime\perp}) = p \, \delta (p + p') \,
  \delta^2 (x^\perp - x^{\prime\perp}) \; ,
\ee
unlike the original expression (\ref{Omega-def}) for $\eta \ne 0$.
Second, both positive and negative longitudinal momenta, $\pm p
\equiv \pm 2 p^+$, contribute in the exponent. Third, all mass
dependence goes away in the LLL, when $\eta^2 \omega^- \to - 4
p^+$.

Thus, also from a functional viewpoint we see that the LLL, which
we have described in terms of spectral flow ($\eta\to0$), has
drastic effects on the Hilbert space of states. This will be
corroborated in the final subsection below.

\subsection{Light-front limit of wave functionals - the Schr\"odinger functional}

We now look at the $\eta\to0$ limit of the Schr\"odinger functional,
\be \label{LFSCHROD}
\fl
  \bra{\varphi_f}\exp - \hat{H}T \ket{\varphi_i} = \int\pathD\phi\,\,\exp
  -\int\limits_0^T\!\ud\xi^0\,L[\phi]\bigg|_{\phi=\varphi^i\
  \mathrm{at}\ \xi^0=0}^{\phi=\varphi^f\ \mathrm{at}\
  \xi^0=T} \equiv \mathcal{N}\exp - S_E(\phi_\mathrm{cl}) \; .
\ee
Again the integral is performed by splitting the field into
orthogonal quantum and classical pieces. Integrating over the
quantum fluctuations yields the prefactor $\mathcal{N}$, which may
be written as the inverse square root of the fluctuation
determinant, $\det \, \hat{\Delta}$. Using the standard heat
kernel identity
\be
  \mathcal{N} = -\frac{1}{2}\log\, \det\, \hat\Delta =
  -\frac{1}{2}\tr\,\log\, \hat\Delta = \frac{1}{2}\int\limits_0^\infty\!
  \frac{\ud s}{s}\,\,\tr\, e^{-s\hat\Delta}, \\
\ee
a regulated determinant $\mathcal{N}_a$ is defined by inserting a
cutoff $a$,
\be
  \mathcal{N}_a\equiv \int\limits_a^\infty\!\frac{\ud s}{s}\,\,\tr\, e^{-s\hat\Delta}\,.
\ee
The classical contribution follows from solving the classical
boundary value problem
\be
  \fl -\partial_\mu \sqrt{g}
  g^{\mu\nu} \partial_\nu \phi - \sqrt{g} m^2 \phi = 0 \; , \quad
  \phi(T, \vc{x})=\varphi^f(\vc{x}) \; , \quad \phi(0, \vc{x}) = \varphi^i(\vc{x}) \;
  .
\ee
The general solution (\ref{cl-field}) now obeys
\be
  \tilde\varphi^i = \tilde\phi_1 + \tilde\phi_2,\qquad
  \tilde\varphi^f = e^{-T\omega^+}\tilde\phi_1 +
  e^{-T\omega^-}\tilde\phi_2.
\ee
and it is straightforward to calculate the corresponding action.
We find, schematically,
\be
  -S_E(\phi_\mathrm{cl}) = \frac{1}{2} \, \varphi^f \ast
  A \ast \varphi^f + \frac{1}{2} \, \varphi^i \ast A \ast \varphi^i -
  \varphi^f \ast B \ast \varphi^i \; ,
\ee
where the asterisks denote convolution integrals. The fields in
the mixed representation depend on $\xi^\perp$ and $p\equiv
k_3\equiv p_-$ and the integral is over all $p$ and $\xi_\perp$.
The kernels are given by
\be\label{AB-def}\fl\eqalign{
  A_p(\xi^\perp,{\xi'}^\perp) &= \int\!\frac{\ud^2 k_\perp}{(2\pi)^2}\,\,
  \frac{\eta^2}{2}(\omega^--\omega^+)\bigg[\frac{1}{e^{T(\omega^+-\omega^-)}-1}+
  \frac{1}{1-e^{T(\omega^--\omega^+)}}\bigg]\,e^{-i k_\perp(\xi-\xi')^\perp}\,, \\
    B_p(\xi^\perp,{\xi'}^\perp) &= \int\!\frac{\ud^2 k_\perp}{(2\pi)^2}\,\,
    \frac{\eta^2}{2}(\omega^--\omega^+)\bigg[
    \frac{e^{-T\omega^-}}{e^{T(\omega^+-\omega^-)}-1} +
    \frac{e^{-T\omega^+}}{1-e^{T(\omega^--\omega^+)}}\bigg]\,e^{-i k_\perp(\xi-\xi')^\perp}\,.
    }
\ee
As $\eta\to0$ we find
\be\eqalign{
  A_p(\xi^\perp,{\xi'}^\perp)&\to\ |p|\,\delta^\perp(\xi-\xi')\,, \\
  B_p(\xi^\perp,{\xi'}^\perp)&\to -p\,
  \theta(p)\int\!\frac{\ud^2 k}{(2\pi)^2}\,\, e^{-T E(p,k_\perp)}
  e^{-ik_\perp(\xi-\xi')^\perp}\,, } \ee
where $E(p,k_\perp) = (k_\perp^2 +m^2)/4p$ is the light-front
energy. The Schr\"odinger functional (\ref{LFSCHROD}) therefore
becomes, in this limit,
\be\label{end}\fl\eqalign{
  \mathcal{N}_a \,\, \exp \Bigg\{ &- \int\limits_0^\infty \!\frac{\ud p}{2\pi}
  \int\! \ud^2\xi \,\, \varphi^f_{-p} (\xi^\perp) \, p \, \varphi^f_p (\xi^\perp)
   -\int\limits_0^\infty \!\frac{\ud p}{2\pi} \int \!\ud^2\xi \,\,
   \varphi^i_{-p} (\xi^\perp) \, p \, \varphi^i_p (\xi^\perp) \\
   & + \int\limits_0^\infty \! \frac{\ud p}{2\pi} \int \!\ud^2\xi \, \ud^2\xi'
   \,\, \varphi^f_{-p} (\xi^\perp) \Bigg[ \int\!\frac{\ud^2 k}{(2\pi)^2} \,\,
   p\, e^{-T E (p, k_\perp)} \, e^{-ik_\perp (\xi-\xi')^\perp} \Bigg] \varphi^i_p(\xi') \Bigg\} \; .
    }
\ee
Comparing with (\ref{LLLVAC}) the terms in the first line are
readily identified with light-front vacuum wave functionals
depending on only initial or final fields $\varphi^i$ and
$\varphi^f$, respectively. Hence, with the fields being stuck at
$x^+ = 0$ and $x^+ = T$, these terms are \textit{non-propagating}.
The final, $T$-dependent, term, on the other hand, \textit{does}
correspond to propagation being precisely the expression for the
anti-holomorphic light-front transition amplitude of appendix~A.
We may thus write the LLL of the Schr{\"o}dinger functional
(\ref{LFSCHROD}) in the compact form
\be
  \bra{\varphi^{f\dagger}} e^{-HT}\ket{\varphi^i} = \mathcal{N}_a\, \Psi_0
  [\varphi^f] \, \bra{\varphi^{f\dagger}} e^{-HT}\ket{\varphi^i}\,
  \Psi_0 [\varphi^i] \; .
\ee
Interestingly, one finds a phenomenon that might be called
bulk-boundary decoupling: the total transition amplitude
decomposes into a product of a bulk and two boundary pieces with
the former to be identified with the first-order, light-front
transition amplitude proper, $\bra{\varphi^{f\dagger}}
e^{-HT}\ket{\varphi^i}$. This is consistent with the observation
that only \textit{half} of the original ($\eta \ne 0$) degrees of
freedom survive outside of the quantisation planes as is manifest
in that the bulk term contains the propagating modes
$\varphi_{p}^{f\dagger}$ and $\varphi^i_{p}$ with only positive
longitudinal momentum, $p>0$. In Fock space language these modes
correspond to creation and annihilation terms corroborating the
interpretation that the propagating $\varphi$'s have indeed become
light-front fields.

The bulk-boundary decoupling, with surface modes $\varphi^f_{-p}$
and $\varphi^i_p$, for $p$ both positive and negative, attached to
the quantisation hypersurfaces, echoes the result of section
\ref{PoisSect} where we saw that the LLL correctly reproduces the
light-front commutation relations only {\it off} the quantisation
surface, i.e.\ in the bulk.

\section{Discussion and conclusions}\label{Conc-sect}

We have discussed limits of several quantum systems in which
second order terms in the action are suppressed. The most striking
feature of these limits is that noncommutativity of configuration
space appears as an \textit{emergent phenomenon} resulting from a
unifying principle, namely \textit{spectral flow}. The limits in
question may then be described in terms of a generic flow
parameter $\lambda$ with $\lambda \to 0$.

We have seen through various applications that it is the
consequential truncation of state space which leads to features of
the first order theory which one expects from a naive treatment of
the classical action. For example, the change of conjugate momenta
to fields rather than their derivatives appears through incomplete
cancellations of mode commutators due to missing states, and the
expected preservation of particle number in nonrelativistic field
theory appears as a restriction on the allowed interactions
controlled by the size of the available energy momentum space.

In the functional picture we have examined the Schr\"odinger and
vacuum functionals. In a Schr\"odinger representation it is only
these two objects which are required to build correlation
functions and S-matrix elements in perturbation theory. We remark
that the form of the boundary terms in the Schr\"odinger
functional, as described by Symanzik \cite{Symanzik:1981wd}, are
key to understanding renormalisation in the Hamiltonian formalism.
This opens up the possibility to interpret the spectral flow
presented here as a \textit{renormalisation group (RG) flow} with
the noncommutativity limit $\lambda \to 0$ corresponding to special RG
fixed points. If this is feasible, the difficult renormalisation
problem of light-front field theory, for instance, might be
attacked from this new vantage point. It is our intention to
address this issue in a future publication.

\subsection*{Acknowledgements}
The authors are grateful to Andreas Wipf for very useful
discussions and A.~I.~ thanks Matthew Daws for helpful
correspondence.

\appendix

\section{Transition amplitudes in the anti-holomorphic representation}

Given a theory with commutator $[\hat\phi(t, \vc{x}),
{\hat\phi}^\dagger (t,\vc{x}')] = \delta^{d} (\vc{x}- \vc{x}')$,
where $\vc{x}$ is shorthand for any set of $d$ dependent
variables, and a Hamiltonian $\hat{H} (\phi^\dagger, \phi)$ one
may describe the abstract space of states $\ket{\Psi, t}$ by wave
functionals $\Psi[ \varphi^\dagger, t] \equiv
\bracket{\varphi^\dagger}{\Psi,t}$ of a complex conjugate field,
on which the operation of $\hat{\phi}^\dagger$ is multiplicative
and $\hat\phi$ acts as a derivative,
\be
  \hat{\phi}^\dagger(0,\vc{x}) \to \varphi^\dagger(\vc{x}),\qquad
  \hat{\phi}(0,\vc{x})\to \frac{\delta}{\delta\varphi^\dagger(\vc{x})}.
\ee
In this representation the time dependence of physical states is
controlled by the Schr\"odinger equation,

\be
  i\frac{\partial}{\partial t}\,\Psi[\varphi^\dagger,t]=
  \hat{H}\bigg(\varphi^\dagger,\frac{\delta}{\delta\varphi^\dagger}
  \bigg)\,\Psi[\varphi^\dagger,t],
\ee
which may be exponentiated to give
\be
  \Psi[\varphi^\dagger,t]=\int\pathD{\varphi'}^\dagger\mathscr{D}\varphi'\,\,
  \bra{\varphi^\dagger}e^{-i\hat{H}t}\ket{\varphi'}\,\Psi[{\varphi'}^\dagger,0].
\ee
The Schr\"odinger functional, $\bra{\varphi^\dagger}e^{-i\hat{H}t}
\ket{\varphi'}$,  may be described by a functional integral
following the usual procedure of discretising the time interval
and inserting complete sets, which are given in this
representation by
\be
  1 = \int\pathD\varphi^\dagger\mathscr{D}\varphi\,\,
  \ket{\varphi} \, \exp \Bigg\{ -\int\!\ud^d
  x\,\varphi^\dagger(\vc{x})\varphi(\vc{x}) \Bigg\} \, \bra{\varphi^\dagger}.
\ee
Note that the measure in this expression is over fields on a
constant time hypersurface. The resulting functional integral is
\be\label{ResFunctInt}\fl\eqalign{
  \bra{\varphi^{\dagger}} e^{-i\hat{H}T} \ket{\varphi'} =
  \int\pathD\phi^\dagger& \mathscr{D}\phi\,\, \exp \Bigg[\frac{1}{2}
  \int\!\ud^d x\,\, \varphi^{\dagger}(\vc{x}) \phi(t,\vc{x}) + \frac{1}{2}
  \int\!\ud^d x \,\, \phi^\dagger(0,\vc{x}) \varphi'(\vc{x}) \\
  &+i\int\limits_{0}^T\!\ud t\!\int\!\ud^d x\,\, \frac{1}{2i}
  (\dot{\phi}^\dagger\phi - \phi^\dagger\dot\phi)- \mathcal{H}(\phi^\dagger,\phi)
  \Bigg]\Bigg|_{\phi(0,\ssvc{x})=\varphi'(\ssvc{x})}^{\phi^\dagger(T,\ssvc{x})
  = \varphi^{\dagger}(\ssvc{x})},
}
\ee
where $\mathcal{H}$ is the Hamiltonian density. This is the
transition amplitude for first order theories derived in the
anti-holomorphic representation using coherent states
\cite{itzykson:1980}. An equivalent expression, which collects all
dependence on the boundary fields into boundary terms in the
action, is
\be\fl\eqalign{
  \bra{\varphi^{\dagger}}e^{-i\hat{H}T}\ket{\varphi'} =
  \int\pathD\phi^\dagger &\mathscr{D}\phi\,\, \exp \Bigg[
  \int\!\ud^d x\,\, \varphi^{\dagger}(\vc{x})\phi(T,\vc{x}) +
  \int\!\ud^d x \,\, \phi^\dagger(0,\vc{x})\varphi'(\vc{x}) \\
  &+i\int\limits_{0}^T\!\ud t\!\int\!\ud^d x\,\,
  \frac{1}{2i}(\dot{\phi}^\dagger\phi-\phi^\dagger\dot\phi)
  -\mathcal{H}(\phi^\dagger,\phi)\Bigg]\Bigg|_{\phi(0,\ssvc{x})=0}^{\phi^\dagger(t,\ssvc{x})=0}
}
\ee
This may be derived either by a rearrangement of terms in the
discretised product, or in the continuum limit using the change of
variables
\be \fl
    \phi^\dagger(t, \vc{x})\rightarrow {\phi}^\dagger(t, \vc{x}) +\theta(t-T) \varphi^{\dagger}(\vc{x}),
    \qquad \phi \rightarrow \phi(t,\vc{x}) + \theta(-t) \varphi'(\vc{x}). \\
\ee
where the new variables obey $\phi^\dagger(T, \vc{x}) =
\phi(0,\vc{x})=0$. This change of variables corresponds to a
separation of degrees of freedom on the boundary, which are fixed
by boundary conditions, and in the bulk, which are integrated
over.

We will finally give the explicit form of the Schr\"odinger
functional for the (Euclidean) light-front field theory of section
4 where we have commutation relations as in (\ref{lc-comm}),
\be\label{Aone}
    \big[\phi_p(x^+,x^\perp),\phi^\dagger_q(x^+,{x'}^\perp)\big]= \delta^\perp(x-x')\,\delta(p-q)\;,
\ee
and Hamiltonian density
\be\label{Atwo}
  \mathcal{H}= \frac{1}{4} \, \phi^\dagger_p(x^+,x^\perp)
  \,(\Delta_\perp^2-m^2) \,\phi_p(x^+,x^\perp) \;.
\ee
The integral in (\ref{ResFunctInt}) is evaluated by first
identifying the solution of the classical equations of motion,
which follow from (\ref{Aone}) and (\ref{Atwo}), subject to the
boundary conditions $\phi^\dagger_p(T,x^\perp) =
\varphi^{\dagger}_p (x^\perp)$, $\phi_p(0,x^\perp)= \varphi'_p
(x^\perp)$. The integration variable may be decomposed into this
field and an orthogonal quantum fluctuation, the integral over the
latter giving a determinant factor which stands in need of
regularisation. One finds,
\be\label{lf-amp}
  \fl \bra{\varphi^\dagger}e^{-\hat H
  T}\ket{\varphi'}=\mathrm{const}. \, \exp \int\limits_0^\infty \! \frac{\ud p}{2\pi} \int \!\ud^2x \,
  \ud^2x' \,\, \varphi^\dagger_p (x^\perp)\, B_p(x^\perp,{x'}^\perp)\, \varphi'_p({x'}^\perp)
\ee
with kernel
\be
  B_p(x^\perp,{x'}^{\perp}) = \int\!\frac{\ud^2 k}{(2\pi)^2}
  \,\, p\, e^{-T E (p, k_\perp)} \, e^{-ik_\perp (x-x')^\perp}\;,
\ee
and light-front energy $E(p,k_\perp)=(k_\perp^2+m^2)/4p$. As
discussed in section 4.5, one may view the light-front theory as
the $\eta\to0$ limit of field theory in the LLL metric. We have
seen that the functional (\ref{lf-amp}) reappears in this limit as
the time dependent (bulk) piece of the LLL Schr\"odinger
functional at $\eta = 0$.

\section{Transition amplitudes in phase space}

The representation given above is appropriate for theories with an
action which is linear in time derivatives and is analogous to the
phase space representation for theories with actions quadratic in
time derivatives. Here, using a real scalar field to illustrate,
it is common to represent states by wave functionals
$\Psi[\phi,t]$. We have the algebra $[\phi(t,\vc{x}),
\pi(t,\vc{y})] = i\delta^d (\vc{x} - \vc{y})$, a Hamiltonian
density $\mathcal{H}(\phi,\pi)$, and complete sets
\be
  1 = \int\pathD\varphi\,\, \ket{\varphi}\bra{\varphi} =
  \int\pathD\pi\,\, \ket{\pi}\bra{\pi}.
\ee
The Schr\"odinger functional may be constructed by discretising
the time interval and inserting complete sets, where we find
\be\fl\eqalign{
  \bra{\varphi_f}e^{-i\hat{H}T}\ket{\varphi_i} =
  \int\pathD\phi \, \mathscr{D}\pi\,\, &\exp \Bigg[\frac{i}{2}\int\!\ud^d x\,\,
  \varphi_f(\vc{x}) \pi(t,\vc{x}) - \frac{i}{2}\int\!\ud^d x\,\, \pi(0, \vc{x}) \varphi_i(\vc{x}) \\
    &+i\int\limits_{0}^T\!\ud t\!\int\!\ud^d x\,\, \frac{1}{2}
    (\pi\dot{\phi}-\dot\pi\phi)-\mathcal{H}(\phi,\pi)\Bigg]\Bigg|_{\phi(0,\ssvc{x})
    =\phi_i(\ssvc{x})}^{\phi(T,\ssvc{x})=\phi_f(\ssvc{x})}.
}
\ee
Equivalently,
\be\fl\eqalign{
    \bra{\varphi_f}e^{-i\hat{H}T}\ket{\varphi_i} = \int \pathD\phi \, \mathscr{D}\pi \, \,
    &\exp \Bigg[ i\int\!\ud^d x\,\,\varphi_f(\vc{x}) \pi(t,\vc{x})-i\int\!\ud^d x\,\,
    \pi(0,\vc{x}) \varphi_i(\vc{x}) \\
    &+i\int\limits_{0}^T\!\ud t\!\int\!\ud^d
    x\,\,\pi\dot{\phi}-\mathcal{H}(\phi,\pi)\Bigg]\Bigg|_{\phi(0,\ssvc{x})=0}^{\phi(T,\ssvc{x})=0}.
} \ee
When the Hamiltonian density is of the form
$\mathcal{H}(\phi,\pi)=\pi^2/2 +V(\phi)$ the momentum integration
may be carried out to leave a configuration space integral over
the exponent of the classical action,
\be\fl\eqalign{
  \bra{\varphi_f}e^{-i\hat{H}T}\ket{\varphi_i} =
  \int\pathD\phi\,\,&\exp \Bigg[ i \int\!\ud^d x\,\,
  \varphi_f(\vc{x}) \dot\phi(t,\vc{x}) - i\int\!\ud^d x\,\,\varphi_i(\vc{x}) \dot\phi(0,\vc{x}) \\
  &+\frac{i}{2}\int\!\ud^d x\,\,\Lambda \varphi^2_f(\vc{x})-
  \frac{i}{2}\int\!\ud^d x\,\,\Lambda\varphi_i^2(\vc{x}) \\
  &+i\int\limits_{0}^T\!\ud t\!\int\!\ud^d x\,\,
  \mathcal{L}(\phi(t,\vc{x})) \Bigg] \Bigg|_{\phi(0,\ssvc{x})=0}^{\phi(T,\ssvc{x})=0}.
} \ee
Here $\Lambda$ is a regularisation of $\delta(0)$ which arises
from the UV (short distance) divergent behaviour of the trivial
field momentum propagator $\delta(t-t')\delta^d(\vc{x}-\vc{x}')$.


\subsection*{References}
\begin{thebibliography}{100}

\bibitem{Snyder:1946qz}
  H.S.~Snyder, \textit{Quantized space-time},
  Phys.\ Rev.\  {\bf 71}, 38 (1947).

\bibitem{Connes:1997cr}
  A.~Connes, M.R.~Douglas and A.S.~Schwarz,
  \textit{Noncommutative geometry and matrix theory: Compactification on
  tori},
  JHEP {\bf 9802}, 003 (1998)
  [arXiv:hep-th/9711162].

\bibitem{Douglas:1997fm}
  M.R.~Douglas and C.M.~Hull,
  \textit{D-branes and the noncommutative torus},
  JHEP {\bf 9802}, 008 (1998)
  [arXiv:hep-th/9711165].

\bibitem{Schomerus:1999ug}
  V.~Schomerus,
  \textit{D-branes and deformation quantization},
  JHEP {\bf 9906}, 030 (1999)
  [arXiv:hep-th/9903205].

\bibitem{Seiberg:1999vs}
  N.~Seiberg and E.~Witten,
  \textit{String theory and noncommutative geometry},
  JHEP {\bf 9909}, 032 (1999)
  [arXiv:hep-th/9908142].

\bibitem{Connes:1994yd}
  A.~Connes, \textit{Noncommutative geometry}, Academic Press,
  1994.

\bibitem{Madore:2000aq}
  J.~Madore,
  \textit{An Introduction To Noncommutative Differential Geometry And Physical
  Applications}, Cambridge University Press, 1999.

\bibitem{Dito:2002dr}
  G.~Dito and D.~Sternheimer,
  \textit{Deformation Quantization: Genesis, Developments and Metamorphoses},
  arXiv:math.qa/0201168.

\bibitem{Douglas:2001ba}
  M.R.~Douglas and N.~A.~Nekrasov,
  \textit{Noncommutative field theory},
  Rev.\ Mod.\ Phys.\  {\bf 73}, 977 (2001)
  [arXiv:hep-th/0106048].

\bibitem{Szabo:2001kg}
  R.J.~Szabo,
  \textit{Quantum field theory on noncommutative spaces},
  Phys.\ Rept.\  {\bf 378}, 207 (2003)
  [arXiv:hep-th/0109162].

\bibitem{Hinchliffe:2002km}
  I.~Hinchliffe, N.~Kersting and Y.L.~Ma,
  \textit{Review of the phenomenology of noncommutative geometry},
  Int.\ J.\ Mod.\ Phys.\  A {\bf 19}, 179 (2004)
  [arXiv:hep-ph/0205040].

\bibitem{Aschieri:2005yw}
  P.~Aschieri, C.~Blohmann, M.~Dimitrijevic, F.~Meyer, P.~Schupp and J.~Wess,
  \textit{A gravity theory on noncommutative spaces},
  Class.\ Quant.\ Grav.\  {\bf 22}, 3511 (2005)
  [arXiv:hep-th/0504183].

\bibitem{Aschieri:2005zs}
  P.~Aschieri, M.~Dimitrijevic, F.~Meyer and J.~Wess,
  \textit{Noncommutative geometry and gravity},
  Class.\ Quant.\ Grav.\  {\bf 23}, 1883 (2006)
  [arXiv:hep-th/0510059].

\bibitem{Calmet:2001na}
  X.~Calmet, B.~Jurco, P.~Schupp, J.~Wess and M.~Wohlgenannt,
  \textit{The standard model on non-commutative space-time},
  Eur.\ Phys.\ J.\  C {\bf 23}, 363 (2002)
  [arXiv:hep-ph/0111115].

\bibitem{Carroll:2001ws}
  S.M.~Carroll, J.~A.~Harvey, V.~A.~Kostelecky, C.~D.~Lane and T.~Okamoto,
  \textit{Noncommutative field theory and Lorentz violation},
  Phys.\ Rev.\ Lett.\  {\bf 87}, 141601 (2001)
  [arXiv:hep-th/0105082].

\bibitem{Susskind:2001fb}
  L.~Susskind,
  \textit{The quantum Hall fluid and non-commutative Chern Simons theory},
  arXiv:hep-th/0101029.

\bibitem{Hellerman:2001rj}
  S.~Hellerman and M.~Van Raamsdonk,
  \textit{Quantum Hall physics equals noncommutative field theory},
  JHEP {\bf 0110}, 039 (2001)
  [arXiv:hep-th/0103179].

\bibitem{Dunne:1989hv}
  G.~V.~Dunne, R.~Jackiw and C.~A.~Trugenberger,
  \textit{Topological (Chern-Simons) Quantum Mechanics},
  Phys.\ Rev.\  D {\bf 41} (1990) 661.

\bibitem{Dunne:1992ew}
  G.~V.~Dunne and R.~Jackiw,
  \textit{`Peierls substitution' and Chern-Simons quantum mechanics},
  Nucl.\ Phys.\ Proc.\ Suppl.\  {\bf 33C}, 114 (1993)
  [arXiv:hep-th/9204057].

\bibitem{Guralnik:2001ax}
  Z.~Guralnik, R.~Jackiw, S.~Y.~Pi and A.~P.~Polychronakos,
  \textit{Testing non-commutative QED, constructing non-commutative MHD},
  Phys.\ Lett.\  B {\bf 517}, 450 (2001)
  [arXiv:hep-th/0106044].

\bibitem{Jackiw:2001dj}
  R.~Jackiw,
  \textit{Physical instances of noncommuting coordinates},
  Nucl.\ Phys.\ Proc.\ Suppl.\  {\bf 108}, 30 (2002)
  [Phys.\ Part.\ Nucl.\  {\bf 33}, S6 (2002\ LNPHA,616,294-304.2003)]
  [arXiv:hep-th/0110057].

\bibitem{Jackiw:2002wd}
  R.~Jackiw,
  \textit{Observations on noncommuting coordinates and on fields depending on
  them},
  Annales Henri Poincare {\bf 4S2}, S913 (2003)
  [arXiv:hep-th/0212146].

\bibitem{landau:1930}
  L.D.~Landau, \textit{Diamagnetismus der {M}etalle}, Z.~Phys.~\textbf{64}, 629 (1930).

\bibitem{landau:1995}
  L.D.~Landau and E.M.~Lifshitz, \textit{The
  Classical Theory of Fields}, Butterworth-Heinemann, 4th revision,
  1995.

\bibitem{johnson:1949}
  M.H.~Johnson and B.A.~Lippmann,
  \textit{Motion in a Constant Magnetic Field},
  Phys.~Rev. \textbf{76}, 828, (1949).


\bibitem{Faddeev:1988qp}
  L.~D.~Faddeev and R.~Jackiw,
  \textit{Hamiltonian Reduction of Unconstrained and Constrained
  Systems},
  Phys.\ Rev.\ Lett.\  {\bf 60}, 1692 (1988).

\bibitem{Peierls:1933}
  R.~Peierls,
  \textit{Zur Theorie des Diamagnetismus von Leitungselektronen},
  Z.~Phys.~\textbf{80}, 763 (1933).

\bibitem{Duval:2000xr}
  C.~Duval and P.~A.~Horvathy,
  \textit{The "Peierls substitution" and the exotic Galilei
  group},
  Phys.\ Lett.\  B {\bf 479}, 284 (2000)
  [arXiv:hep-th/0002233].

\bibitem{Ambjorn:2000yr}
  J.~Ambjorn, Y.~M.~Makeenko, G.~W.~Semenoff and R.~J.~Szabo,
  \textit{String theory in electromagnetic fields},
  JHEP {\bf 0302}, 026 (2003)
  [arXiv:hep-th/0012092].

\bibitem{Chu:1998qz}
  C.~S.~Chu and P.~M.~Ho,
  \textit{Noncommutative open string and D-brane},
  Nucl.\ Phys.\  B {\bf 550}, 151 (1999)
  [arXiv:hep-th/9812219].

\bibitem{Szabo:2004ic}
  R.~J.~Szabo,
  \textit{Magnetic backgrounds and noncommutative field theory},
  Int.\ J.\ Mod.\ Phys.\  A {\bf 19}, 1837 (2004)
  [arXiv:physics/0401142].

\bibitem{zee:2003}
  A.~Zee,
  \textit{Quantum Field Theory in a Nutshell},
  Princeton University Press, 2003.

\bibitem{Beg:1984yh}
  M.~A.~B.~Beg and R.~C.~Furlong,
  \textit{The $\lambda \phi^4$ Theory in the Nonrelativistic Limit},
  Phys.\ Rev.\  D {\bf 31}, 1370 (1985).

\bibitem{Tamm:1945qv}
  I.~Tamm,
  \textit{Relativistic Interaction Of Elementary Particles},
  J.\ Phys.\ (USSR) {\bf 9}, 449 (1945).

\bibitem{Dancoff:1950ud}
  S.~M.~Dancoff,
  \textit{Nonadiabatic meson theory of nuclear forces},
  Phys.\ Rev.\  {\bf 78}, 382 (1950).

\bibitem{Dirac:1949cp}
  P.~A.~M.~Dirac,
  \textit{Forms Of Relativistic Dynamics},
  Rev.\ Mod.\ Phys.\  {\bf 21}, 392 (1949).

\bibitem{Stueckelberg:1951}
  E.C.G.~Stueckelberg,
  \textit{Relativistic Quantum Theory for Finite Time Intervals},
  Phys.~Rev.~\textbf{81}, 130 (1951).

\bibitem{Bogolubov:1959}
  N.N.~Bogolubov and D.V.~Shirkov,
  \textit{Introduction to the theory of quantized fields},
  Interscience, 1959.

\bibitem{Symanzik:1981wd}
  K.~Symanzik,
  \textit{Schr{\"o}dinger Representation And Casimir Effect In Renormalizable Quantum Field Theory},
  Nucl.\ Phys.\  B {\bf 190}, 1 (1981).

\bibitem{Namyslowski:1985zq}
  J.~M.~Namyslowski,
  \textit{Light Cone Perturbation Theory And Its Application To Different Fields},
  Prog.\ Part.\ Nucl.\ Phys.\  {\bf 14}, 49 (1985).

\bibitem{Burkardt:1995ct}
  M.~Burkardt, \textit{Light front quantization},
  Adv.\ Nucl.\ Phys.\  {\bf 23}, 1 (1996)
  [arXiv:hep-ph/9505259].

\bibitem{Brodsky:1997de}
  S.~J.~Brodsky, H.~C.~Pauli and S.~S.~Pinsky,
  \textit{Quantum chromodynamics and other field theories on the light
  cone},
  Phys.\ Rept.\  {\bf 301}, 299 (1998)
  [arXiv:hep-ph/9705477].

\bibitem{Heinzl:2000ht}
  T.~Heinzl,
  \textit{Light-cone quantization: Foundations and applications},
  Lect.\ Notes Phys.\  {\bf 572}, 55 (2001)
  [arXiv:hep-th/0008096].

\bibitem{Perry:1990mz}
  R.~J.~Perry, A.~Harindranath and K.~G.~Wilson,
  \textit{Light front Tamm-Dancoff field theory},
  Phys.\ Rev.\ Lett.\  {\bf 65}, 2959 (1990).

\bibitem{Perry:1991ny}
  R.~J.~Perry and A.~Harindranath,
  \textit{Renormalization in the light front Tamm-Dancoff approach to field
  theory},
  Phys.\ Rev.\  D {\bf 43}, 4051 (1991).

\bibitem{Weinberg:1966jm}
  S.~Weinberg,
  \textit{Dynamics at infinite momentum},
  Phys.\ Rev.\  {\bf 150}, 1313 (1966).

\bibitem{Heinzl:2006dw}
  T.~Heinzl,
  \textit{Alternative approach to light-front perturbation theory},
  Phys.\ Rev.\  D {\bf 75}, 025013 (2007)
  [arXiv:hep-ph/0610293].

\bibitem{Susskind:1967rg}
  L.~Susskind,
  \textit{Model of self-induced strong interactions},
  Phys.\ Rev.\  {\bf 165}, 1535 (1968).

\bibitem{Bardakci:1969dv}
  K.~Bardakci and M.~B.~Halpern,
  \textit{Theories at infinite momentum},
  Phys.\ Rev.\  {\bf 176}, 1686 (1968).

\bibitem{Chen:1971yg}
  T.~W.~Chen,
  \textit{Almost-infinite-momentum frame and high-energy scattering
  processes},
  Phys.\ Rev.\  D {\bf 3}, 2257 (1971).

\bibitem{Frishman:1976vi}
  Y.~Frishman, C.~T.~Sachrajda, H.~D.~I.~Abarbanel and R.~Blankenbecler,
  \textit{A Novel Inconsistency In Two-Dimensional Gauge
  Theories},
  Phys.\ Rev.\  D {\bf 15}, 2275 (1977).

\bibitem{Hornbostel:1991qj}
  K.~Hornbostel,
  \textit{Nontrivial Vacua From Equal Time To The Light Cone},
  Phys.\ Rev.\  D {\bf 45}, 3781 (1992).

\bibitem{Prokhvatilov:1989eq}
  E.~V.~Prokhvatilov and V.~A.~Franke,
  \textit{Limiting transition of light-front coordinates in field theory and the  QCD
  Hamiltonian},
  Sov.\ J.\ Nucl.\ Phys.\  {\bf 49}, 688 (1989)
  [Yad.\ Fiz.\  {\bf 49}, 1109 (1989)].

\bibitem{Lenz:1991sa}
  F.~Lenz, M.~Thies, K.~Yazaki and S.~Levit,
  \textit{Hamiltonian formulation of two-dimensional gauge theories on the light
  cone},
  Annals Phys.\  {\bf 208}, 1 (1991).

\bibitem{Prokhvatilov:1994dm}
  E.~V.~Prokhvatilov, H.~W.~L.~Naus and H.~J.~Pirner,
  \textit{Effective light-front quantization of scalar field theories and
  two-dimensional electrodynamics},
  Phys.\ Rev.\  D {\bf 51}, 2933 (1995)
  [arXiv:hep-ph/9406275].

\bibitem{Naus:1997zg}
  H.~W.~L.~Naus, H.~J.~Pirner, T.~J.~Fields and J.~P.~Vary,
  \textit{QCD near the light cone},
  Phys.\ Rev.\  D {\bf 56}, 8062 (1997)
  [arXiv:hep-th/9704135].

\bibitem{Ilgenfritz:2006ir}
  E.~M.~Ilgenfritz, S.~A.~Paston, H.~J.~Pirner, E.~V.~Prokhvatilov and V.~A.~Franke,
  \textit{Quantum Fields on the Light Front, Formulation in Coordinates close to the
  Light Front, Lattice Approximation},
  Theor.\ Math.\ Phys.\  {\bf 148}, 948 (2006)
  [Teor.\ Mat.\ Fiz.\  {\bf 148}, 89 (2006)]
  [arXiv:hep-th/0610020].

\bibitem{Alves:2002tx}
  V.~S.~Alves, A.~Das and S.~Perez,
  \textit{Light-front field theories at finite temperature},
  Phys.\ Rev.\  D {\bf 66}, 125008 (2002)
  [arXiv:hep-th/0209036].

\bibitem{Das:2004je}
  A.~Das and S.~Perez,
  \textit{Quantization in a general light-front frame},
  Phys.\ Rev.\  D {\bf 70}, 065006 (2004)
  [arXiv:hep-th/0404200].

\bibitem{Hellerman:1997yu}
  S.~Hellerman and J.~Polchinski, \textit{Compactification in the lightlike
  limit},
  Phys.\ Rev.\  D {\bf 59}, 125002 (1999)
  [arXiv:hep-th/9711037].

\bibitem{Yamawaki:1998cy}
  K.~Yamawaki,
  \textit{Zero-mode problem on the light front},
  arXiv:hep-th/9802037.

\bibitem{Heinzl:2003jy}
  T.~Heinzl,
  \textit{Light-cone zero modes revisited},
  in: {P}roceedings of the International Workshop on
  Light-Cone Physics: Hadrons and Beyond, IPPP/03/71,
  Durham, 2003; S.~Dalley, ed.
  arXiv:hep-th/0310165.

\bibitem{Jaramillo:1998ff}
  A.~Jaramillo and P.~Mansfield,
  \textit{Finite VEVs from a large distance vacuum wave functional},
  Int.\ J.\ Mod.\ Phys.\  A {\bf 15}, 581 (2000)
  [arXiv:hep-th/9808067].

\bibitem{Birmingham:1986gj}
  D.~Birmingham and C.~G.~Torre,
  \textit{Functional Integral Construction Of The BRST Invariant String Ground State},
  Class.\ Quant.\ Grav.\  {\bf 4}, 1149 (1987).

\bibitem{Hartle:1983ai}
  J.~B.~Hartle and S.~W.~Hawking,
  \textit{Wave Function Of The Universe},
  Phys.\ Rev.\  D {\bf 28}, 2960 (1983).

\bibitem{itzykson:1980}
  C.~Itzykson and J.B.~Zuber, \textit{Quantum Field Theory},
  McGraw-Hill, 1980.

\end{thebibliography}
\end{document}